\documentclass[10pt]{article}
\usepackage{amsmath}
\usepackage{mathrsfs}
\usepackage{amsfonts}
\usepackage{amssymb}
\usepackage{graphicx, times}
\usepackage{subfig}
\usepackage{amsthm}
\usepackage{bm}
\usepackage[left=2.5cm, right=3cm, top=2cm]{geometry}
\usepackage[section]{placeins}  
\newtheorem{The}{Theorem}[section]
\newtheorem{Def}{Definition}[section]

\newtheorem{conj}{Conjecture}

\begin{document}
\begin{center}
{\LARGE {\bf A new fractional order chaotic dynamical system and its synchronization using optimal control}}
\vskip 1cm
{\Large  Madhuri Patil$^1$, Sachin Bhalekar$^{1,2}$}\\
\textit{$^1$Department of Mathematics, Shivaji University, Kolhapur - 416004, India, $^2$ School of Mathematics and Statistics, University of Hyderabad, Hyderabad, India, Email:madhuripatil4246@gmail.com (Madhuri Patil), sachin.math@yahoo.co.in, sbb\_maths@unishivaji.ac.in (Sachin Bhalekar)}\\
\end{center}
\begin{abstract}
In this work, we introduce a new three-dimensional chaotic differential dynamical system. We find equilibrium points of this system and provide the stability conditions for various fractional orders. Numerical simulations will be used to investigate the chaos in the proposed system. A simple linear control will be used to control the chaotic oscillations. Further, we propose an optimal control which is based on the fractional order of the system and use it to synchronize new chaotic system.
\end{abstract}

\section{Introduction}
The differential equation is a prime tool used by Scientists in modeling various natural phenomena. To make it more realistic, the generalized operator viz. fractional derivative \cite{Podlubny,Diethelm} is introduced by the researchers. The order in fractional derivative can be any real or complex number, a function of time or may be distributed over some interval. This flexible order makes fractional derivative more suitable to model the intermediate processes and the memory properties. The fractional differential equations (FDE) have applications in Bioengineering \cite{Magin}, Viscoelasticity \cite{Mainardi}, Control theory \cite{Baleanu}, and so on. The analysis of FDEs is presented in \cite{Matignon1,Tavazoei-Haeri,Samko,Miller,Zhang,Gejji1,Babakhani}. The difficult task of numerical solutions of FDEs is handled in \cite{DFF, V. Gejji 3}.

\par The signals generated by a higher-order deterministic nonlinear system which are aperiodic for all the time and depend sensitively on initial conditions are termed as ``chaotic" \cite{Devaney,Alligood}. The chaos can occur in a nonlinear autonomous differential dynamical system of order three or more, a delay differential equation and a discrete map. The most celebrated examples of chaos are the Lorenz system  \cite{Lorenz} and the logistic map \cite{May}. Few other examples include the systems viz. Chen \cite{Chen}, Chua \cite{Matsumoto}, Lu  \cite{Lu}, Rossler \cite{Rossler}, Bhalekar-Gejji(BG) \cite{BG}, Modified BG \cite{Singh-Roy}, Proto BG  \cite{Aqeel}, Pehlivan \cite{Sundarpandian}, and so on.
\par The fractional order counterparts of these classical systems also produce chaotic signals for certain values. For a commensurate order case, it is observed that the system remains chaotic up to some threshold value of fractional order and then becomes stable. Few examples of fractional order chaotic systems are described in \cite{Grigorenko, Hartley, fractional-Liu, Bhalekar-Baleanu, Yadav, Faieghi, I.Petras}.
\par Chaos in fractional ordered delayed systems is analyzed in \cite{Bhalekar liu-delay, DDE2, DDE3, DDE4, DDE5}.

\par The paper is organized as follows:\\
The basics are described in Section \ref{section2}. We propose a new chaotic system and present the stability, bifurcation and chaos in Section \ref{section3}. Sections \ref{section5} and \ref{section6} deal with the chaos control and synchronization, respectively. The comments on the incommensurate order case are presented in Section \ref{section7}. The conclusions are summarized in Section \ref{section8}.  

\section{Preliminaries} \label{section2}
This section deals with basic definitions and results given in the literature \cite{Podlubny,Diethelm,Samko,Das}. Throughout this section, we take $n\in\mathbb{N}$.
\begin{Def} \label{Def 2.1}
Let $\alpha\ge0$ \,\, ($\alpha\in\mathbb{R}$). Then Riemann-Liouville (\text RL) fractional integral of function $f\in C[0,b]$, $t>0$ of order `$\alpha$' is defined as,
\begin{equation}
{}_0\mathrm{I}_t^\alpha f(t)=
\frac{1}{\Gamma{(\alpha)}}\int_0^t (t-\tau)^{\alpha-1}f(\tau)\,\mathrm{d}\tau. \label{2.1}
\end{equation}
\end{Def}
\begin{Def}
The Riemann-Liouville (\text RL) fractional derivative of order $\alpha>0$ of function $f\in C[0,b]$,\, $t>0$ is defined as, 
\begin{equation}
{}^{RL}_0\mathrm{D}_t^\alpha f(t)=
\begin{cases}
\frac{1}{\Gamma{(n-\alpha)}}\frac{d^n}{dt^n}\int_0^t (t-\tau)^{n-\alpha-1}f(\tau)\,\mathrm{d}\tau, & \mathrm{if}\,\, n-1<\alpha< n\\
\frac{d^n}{dt^n}f(t), & \mathrm{if}\, \alpha=n.
\end{cases} 
\end{equation}
\end{Def}
\begin{Def}\label{Def 2.2}
The Caputo fractional derivative of order $\alpha>0$, $n-1<\alpha \le n$ is defined for $f\in C^n[0,b]$,\, $t>0$ \,as,
\begin{equation}
{}_0^{C}\mathrm{D}_t^\alpha f(t)=
\begin{cases}
\frac{1}{\Gamma{(n-\alpha)}}\int_0^t (t-\tau)^{n-\alpha-1}f^{(n)}(\tau)\,\mathrm{d}\tau, & \mathrm{if}\,\, n-1<\alpha< n\\
\frac{d^n}{dt^n}f(t), & \mathrm{if}\,\, \alpha=n.
\end{cases}\label{2.2}
\end{equation}
 
\end{Def}
\begin{Def} \label{Def 2.3}
The one-parameter Mittag-Leffler function is defined as,
\begin{equation}
E_\alpha(z)=\sum_{k=0}^\infty \frac{z^k}{\Gamma(\alpha k+1)}\, ,\qquad z\in\mathbb{C}, \,\,(\alpha>0).\label{2.3}
\end{equation}
The two-parameter Mittag-Leffler function is defined as,
\begin{equation}
E_{\alpha,\beta}(z)=\sum_{k=0}^\infty \frac{z^k}{\Gamma(\alpha k+\beta)}\, ,\qquad z\in\mathbb{C}, \,\,(\alpha>0,\,\beta>0).\label{2.4}
\end{equation}
\end{Def}

\noindent{\bf Properties}\\
(i) Let $n-1< \alpha \le n$ and $\beta\ge0$\\
\quad ${}_0^{C}\mathrm{D}_t^\alpha t^\beta=
\begin{cases}
\frac{\Gamma(\beta+1)}{\Gamma(-\alpha+\beta+1)}t^{\beta-\alpha} ,\, \mathrm{if}\, \beta>n-1, \, \beta\in\mathbb{R} \\
0\qquad \qquad \quad \quad\, ,\, \mathrm{if}\, \beta\in \{0,1,2,\dots,n-1\}.
\end{cases}$\\
(ii) ${}_0^{C}\mathrm{D}_t^\alpha {}_0\mathrm{I}_t^\beta f(t)=
\begin{cases} 
{}_0\mathrm{I}_t^{\beta-\alpha }f(t), \quad \mathrm{if} \,   \beta>\alpha\\
f(t) , \qquad \quad \,\,\,  \mathrm{if} \, \beta=\alpha\\
{}_0^{C}\mathrm{D}_t^{\alpha-\beta} f(t), \,\, \mathrm{if} \, \alpha> \beta.
\end{cases}$\\
(iii)${}_0^{C}\mathrm{D}_t^\alpha c=0$, where $c$ is a constant.\\
(iv) ${}_0^{RL}\mathrm{D}_t^\alpha c=\frac{c \,t^{-\alpha}}{\Gamma(1-\alpha)}$, where $c$ is a constant.
\begin{The} \label{Thm 2.1}
\cite{Luchko} Solution of homogeneous fractional order differential equation
\begin{equation}
{}_0^C\mathrm{D}_t^\alpha x(t)+\lambda x(t)=0, \qquad 0<\alpha<1 \label{2.8}
\end{equation} 
is given by,
\begin{equation}
x(t)=x(0)E_\alpha(-\lambda t^\alpha).\label{2.9}
\end{equation}
\end{The}

\noindent{\bf \large Stability Analysis}\\

Consider the following fractional order system,
\begin{equation}
\begin{split}
{}_0^C\mathrm{D}_t^{\alpha_1}x_1 &=f_1(x_1, x_2, \dots, x_n),\\
{}_0^C\mathrm{D}_t^{\alpha_2}x_2 &=f_2(x_1, x_2, \dots, x_n),\\
&\quad\vdots\\
{}_0^C\mathrm{D}_t^{\alpha_n}x_n &=f_n(x_1, x_2, \dots, x_n)
\end{split} \label{5.2}
\end{equation}
where $0<\alpha_i<1$ are fractional orders. If $\alpha_1=\alpha_2=\cdots=\alpha_n$, then the system (\ref{5.2}) is called commensurate order system, otherwise incommensurate order system.
\par A point $E=(x_1^*, x_2^*,\dots, x_n^*)$ is called an equilibrium point of the system (\ref{5.2}) if 
$$
f_i(E)=f_i(x_1^*, x_2^*,\dots, x_n^*)=0, \quad\mathrm{for} \,\, \mathrm{each}\quad i=1,2,\dots,n.
$$

\noindent{\bf (a) Commensurate order system:} \\
\begin{The}\cite{Tavazoei, Matignon}
	Consider $\alpha=\alpha_1=\alpha_2=\cdots=\alpha_n$ in (\ref{5.2}). An equilibrium point $E$ of the system (\ref{5.2}) is locally asymptotically stable if all the eigenvalues of the Jacobian matrix
	\begin{equation}
	J=\begin{bmatrix}
	\frac{\partial  f_1}{\partial  x_1} & \frac{\partial  f_1}{\partial  x_2} & \cdots & \frac{\partial  f_1}{\partial  x_n}\\
	\frac{\partial  f_2}{\partial  x_1} & \frac{\partial  f_2}{\partial  x_2} & \cdots & \frac{\partial  f_2}{\partial  x_n}\\
	\vdots & \vdots & \vdots & \vdots\\
	\frac{\partial  f_n}{\partial  x_1} & \frac{\partial  f_n}{\partial  x_2} & \cdots & \frac{\partial  f_n}{\partial  x_n}
	\end{bmatrix}\label{5.3}
	\end{equation}
	evaluated at $E=(x_1^*,x_2^*, \dots, x_n^*)$ satisfy the following condition
	$$ \left|\mathrm{arg}(\mathrm{Eig}(J|_E))\right|>\frac{\alpha \pi}{2}. $$
\end{The}

\noindent{\bf (b) Incommensurate order system:}\\
\begin{The}\cite{Tavazoei-Haeri}
	Consider the incommensurate fractional ordered dynamical system given by (\ref{5.2}). Let $\alpha_i=\frac{v_i}{u_i}$, gcd$(u_i,v_i)=1$, $u_i$, $v_i$ be positive integers. Define $M$ to be the least common multiple of $u_i$'s. \\
	Define,
	\begin{equation}
	\Delta (\lambda)=\mathrm{diag} \left(\left[\lambda^{M \alpha_1}, \lambda^{M \alpha_2}, \dots, \lambda^{M \alpha_n}\right]\right)-J
	\end{equation} \label{5.4}
\end{The}
where, $J$ is the Jacobian matrix as defined in (\ref{5.3}) evaluated at point $E$. If all the roots $\lambda$'s of $\mathrm{det}(\Delta (\lambda))=0$ satisfy $|\mathrm{arg}(\lambda)|>\frac{\alpha \pi}{2}$, then $E$ is locally asymptotically stable. This condition is equivalent to the following inequality
\begin{equation}
\frac{\pi}{2 M}-\min_{\substack{i}} |\mathrm{arg}(\lambda_i)|<0.
\end{equation}\label{5.5}
The term $\frac{\pi}{2 M}-\min_{\substack{i}} |\mathrm{arg}(\lambda_i)|$ is called as the instability measure for equilibrium points in fractional order systems (IMFOS). Hence, a necessary condition \cite{Tavazoei-Haeri} for fractional order system (\ref{5.2}) to exhibit chaotic attractor is
\begin{equation}
\mathrm{IMFOS} \ge 0. \label{5.6}
\end{equation}
Note that, the condition (\ref{5.6}) is not sufficient \cite{Tavazoei-Haeri,fractional-Liu} for chaos to exist.

\section{New chaotic system} \label{section3}
We propose the following chaotic dynamical system,
\begin{equation}
\begin{split}
\dot{x}&=a x-y^2,\\
\dot{y}&=b y-z+d xz,\\
\dot{z}&=-g z+4xy-h x^2,
\end{split}\label{5.1.1}
\end{equation}
where $a$, $b$, $d$, $g$, $h$ $\in \mathbb{R}$ are  parameters. When $a=-1$, $b=2.5$, $d=-5$, $g=5.5$ and $h=-0.2$, system (\ref{5.1.1}) shows chaotic behavior.\\
If $X_*=(x_*,y_*,z_*)$ is an equilibrium point of (\ref{5.1.1}), then the Jacobian $J(X_*)$ of this system at $X_*$ is given by 
$$ J(X_*)= \begin{bmatrix}
 a & -2 y_* & 0 \\
d z_* & b & -1 + d x_* \\
 -2 h x_* + 4 y_* & 4 x_* & -g
\end{bmatrix}. $$

\subsection{Bifurcation Analysis}
\begin{figure*}
	\begin{tabular}{c c}
		\subfloat[$a=-4$ ]{\includegraphics[width=0.55\textwidth]{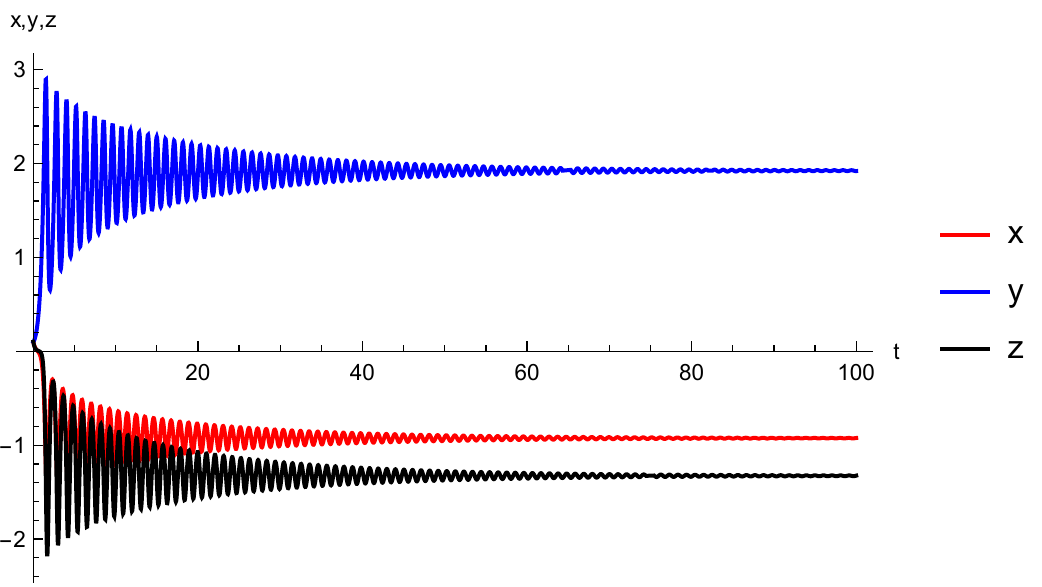}} 
		& 
		\subfloat[$a=-3.5$  ]{\includegraphics[width=0.2\textwidth]{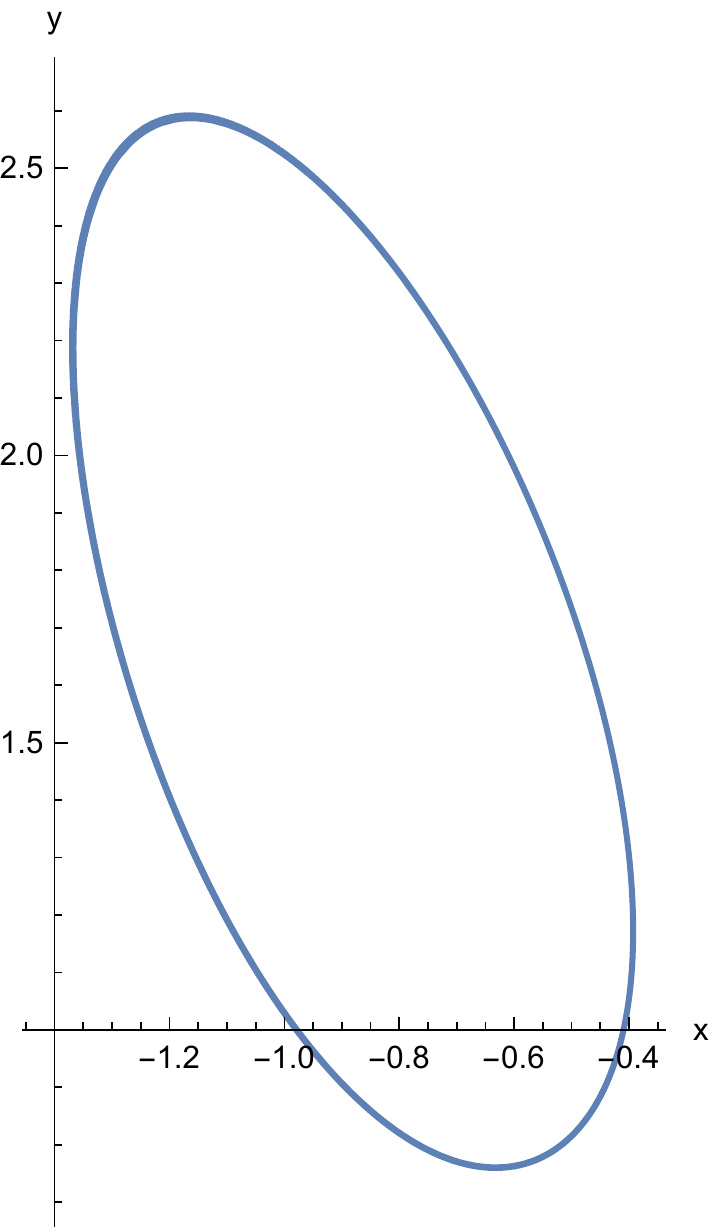}}
		\\
	\subfloat[$a=-2.8$ ]{\includegraphics[width=0.15\textwidth]{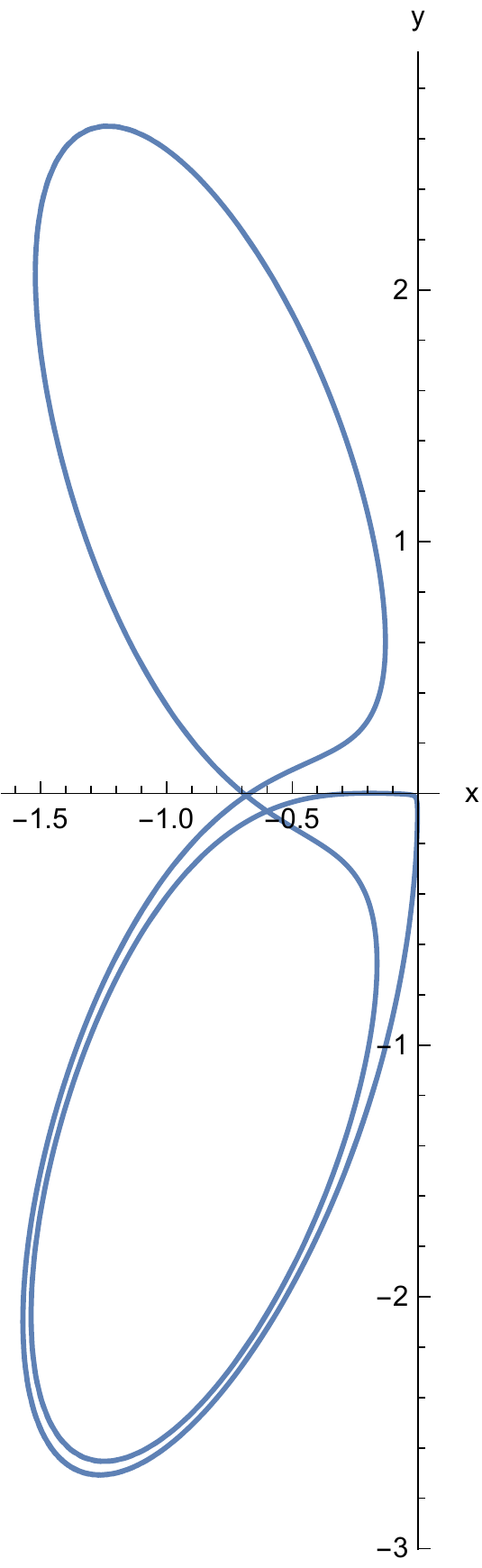}}	
		& 
		\subfloat[$a=-2.6$]{\includegraphics[width=0.15\textwidth]{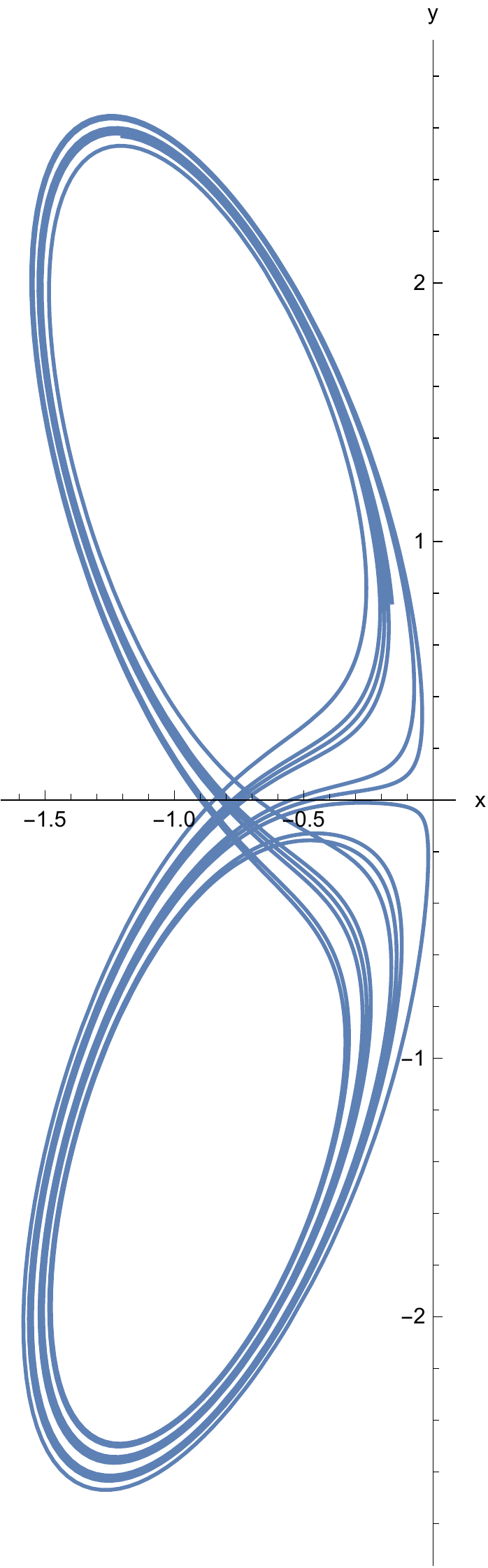}} 
		\\
		\subfloat[$a=-2$]{\includegraphics[width=0.15\textwidth]{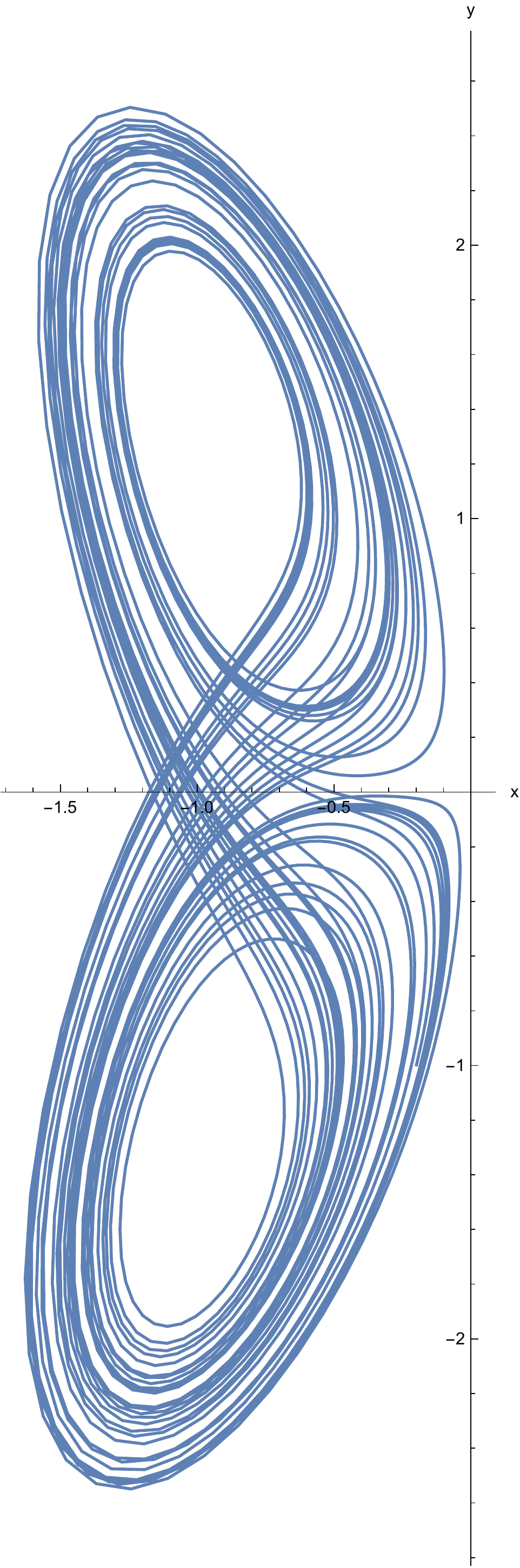}} 
		& 
		\subfloat[Bifurcation with parameter $a$]{\includegraphics[width=0.5\textwidth]{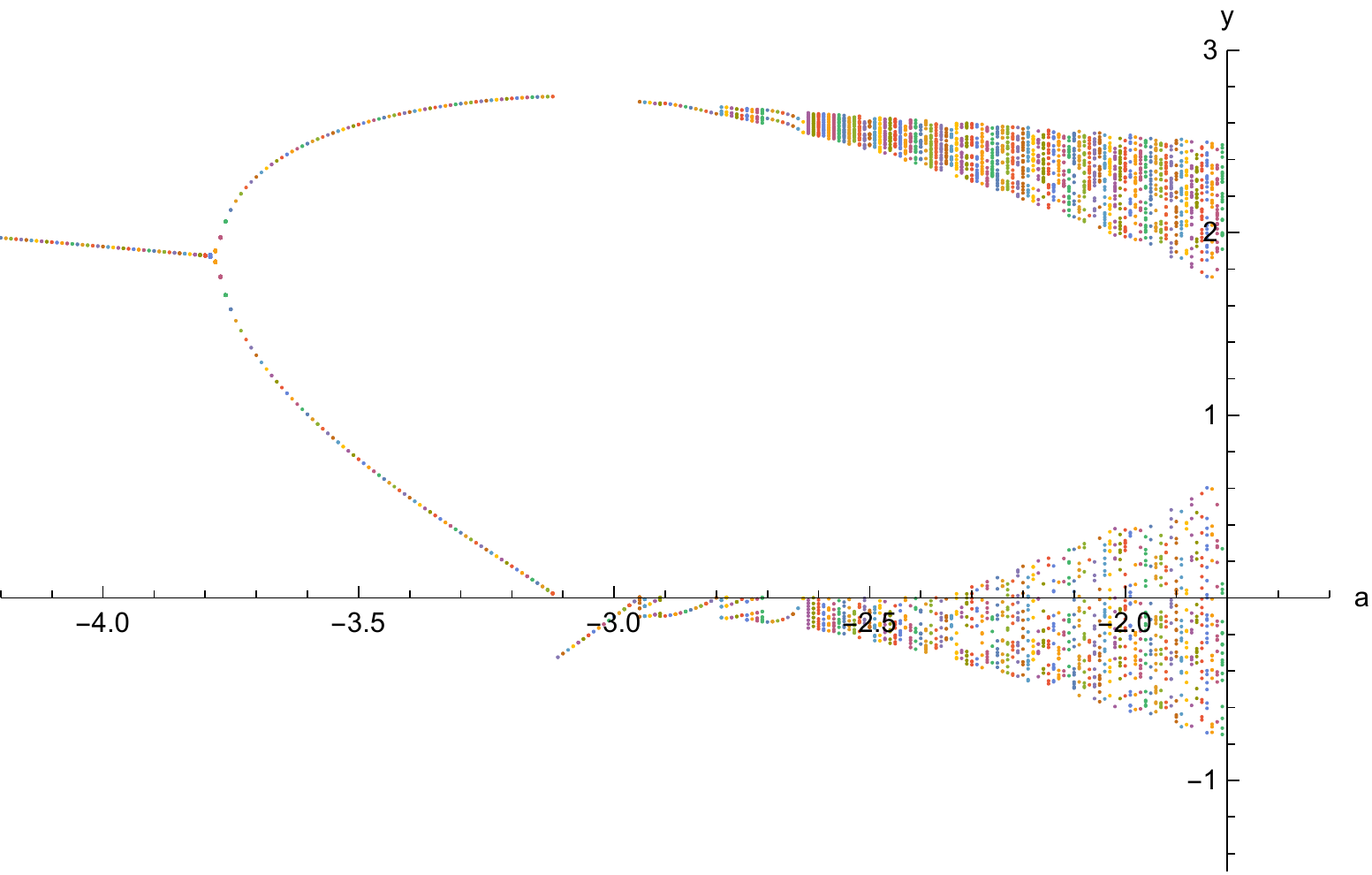}} 
	\end{tabular}
	\caption{Bifurcation analysis for parameter $a$}
	\label{Figa}
\end{figure*}

\begin{figure*}
	\begin{tabular}{c c}
		\subfloat[$b=1.6$ ]{\includegraphics[width=0.42\textwidth]{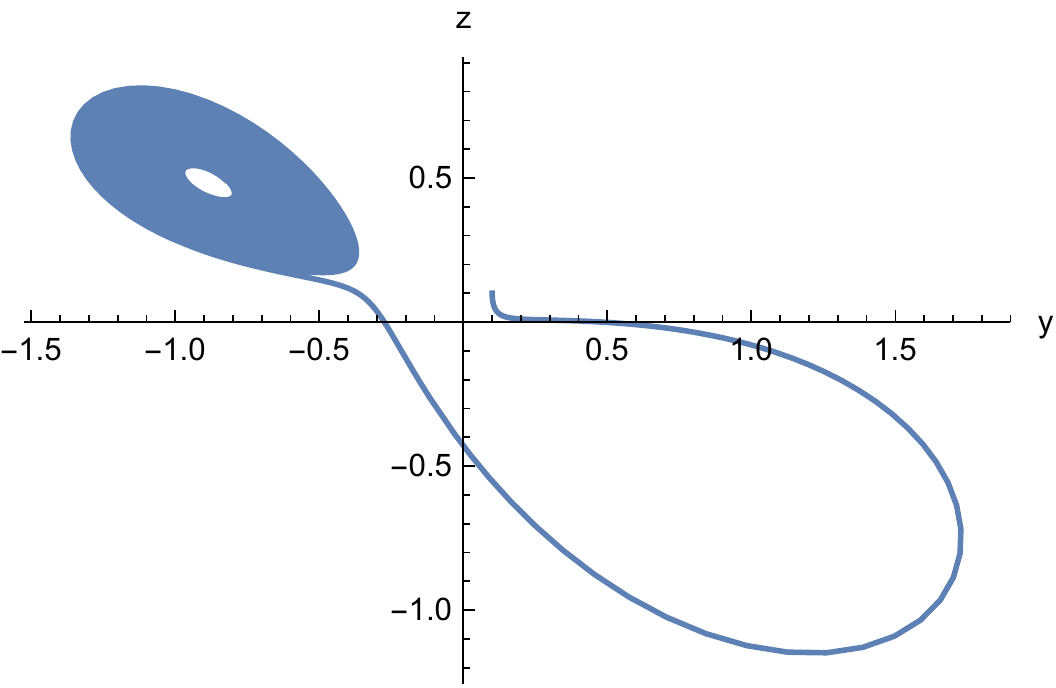}} 
		& 
		\subfloat[$b=1.8$  ]{\includegraphics[width=0.42\textwidth]{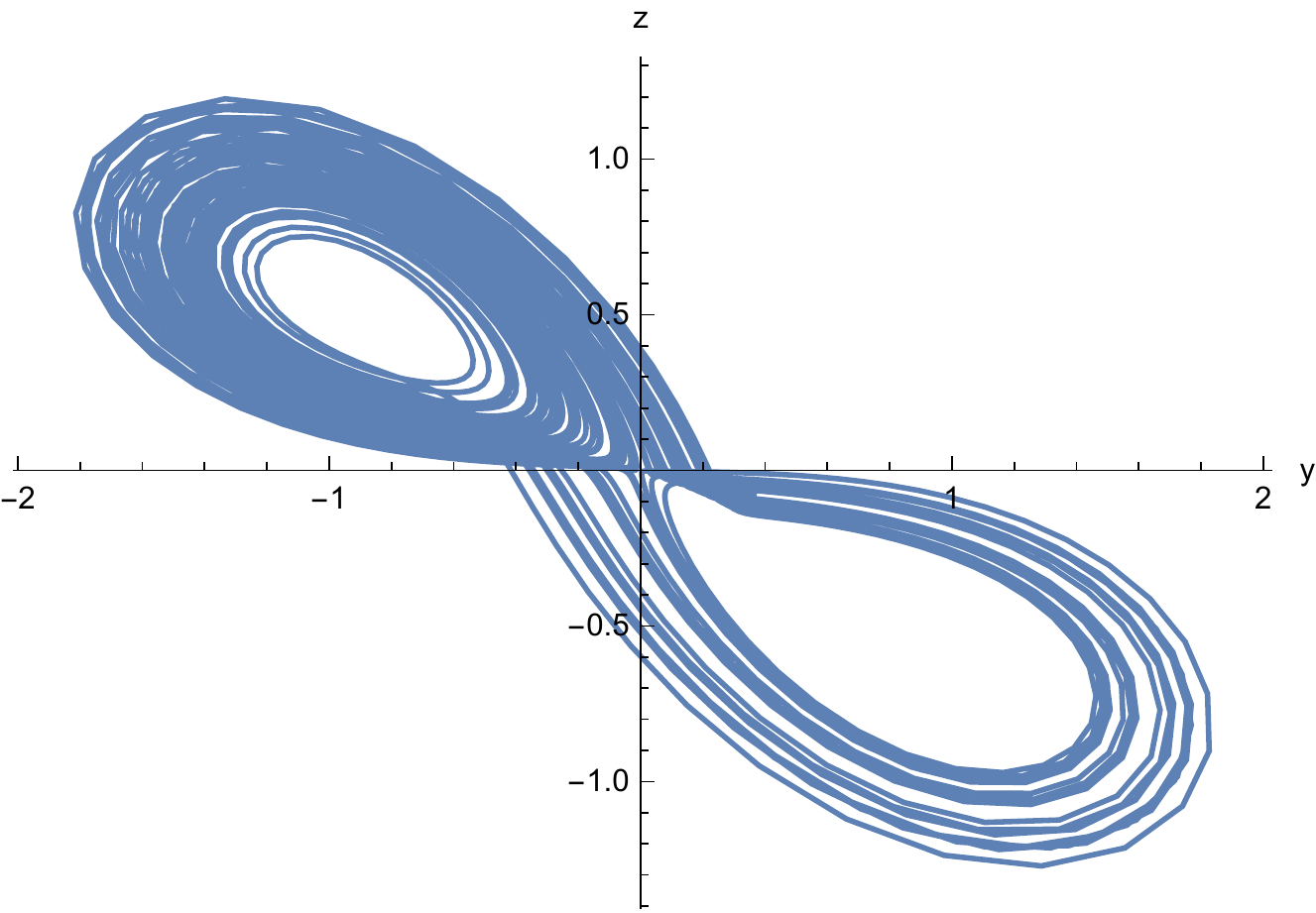}}
	    \\
		\subfloat[$b=2.5$ ]{\includegraphics[width=0.39\textwidth]{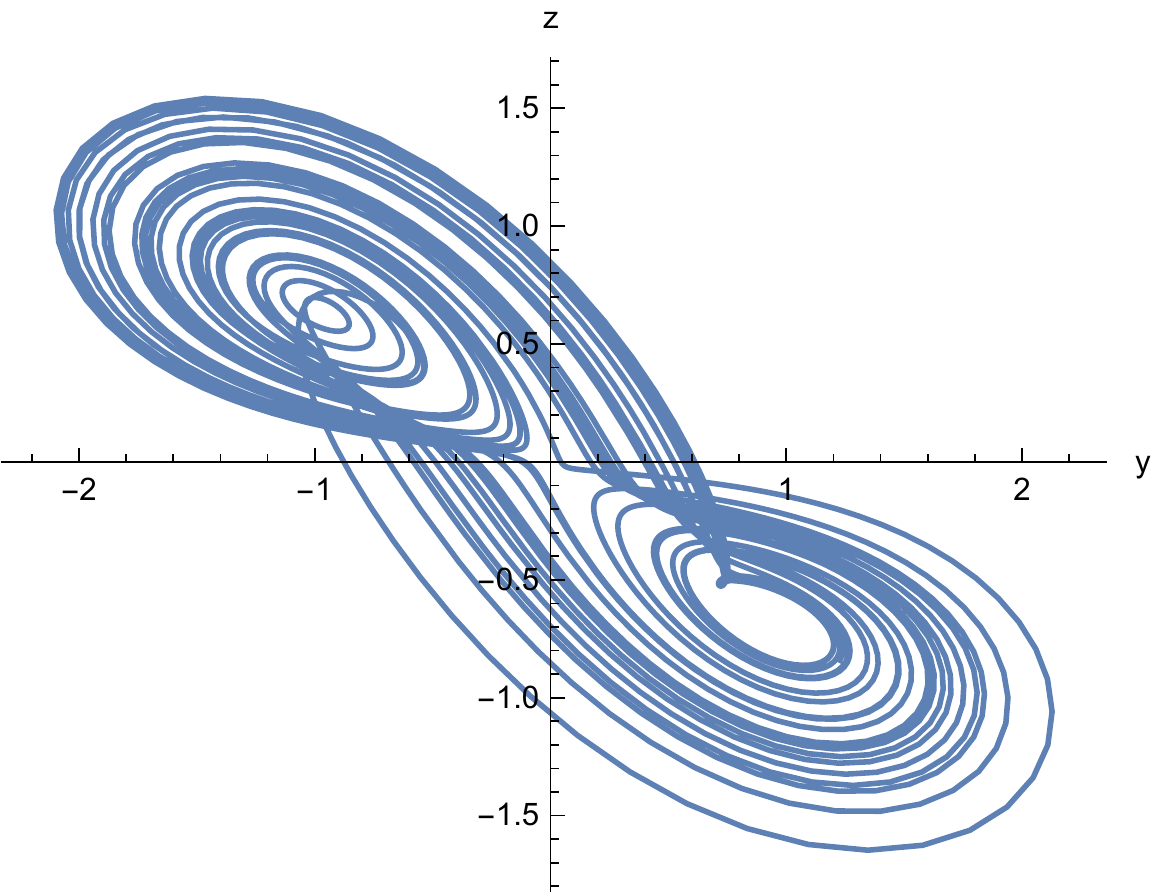}}	
		& 
		\subfloat[$b=4.35$]{\includegraphics[width=0.35\textwidth]{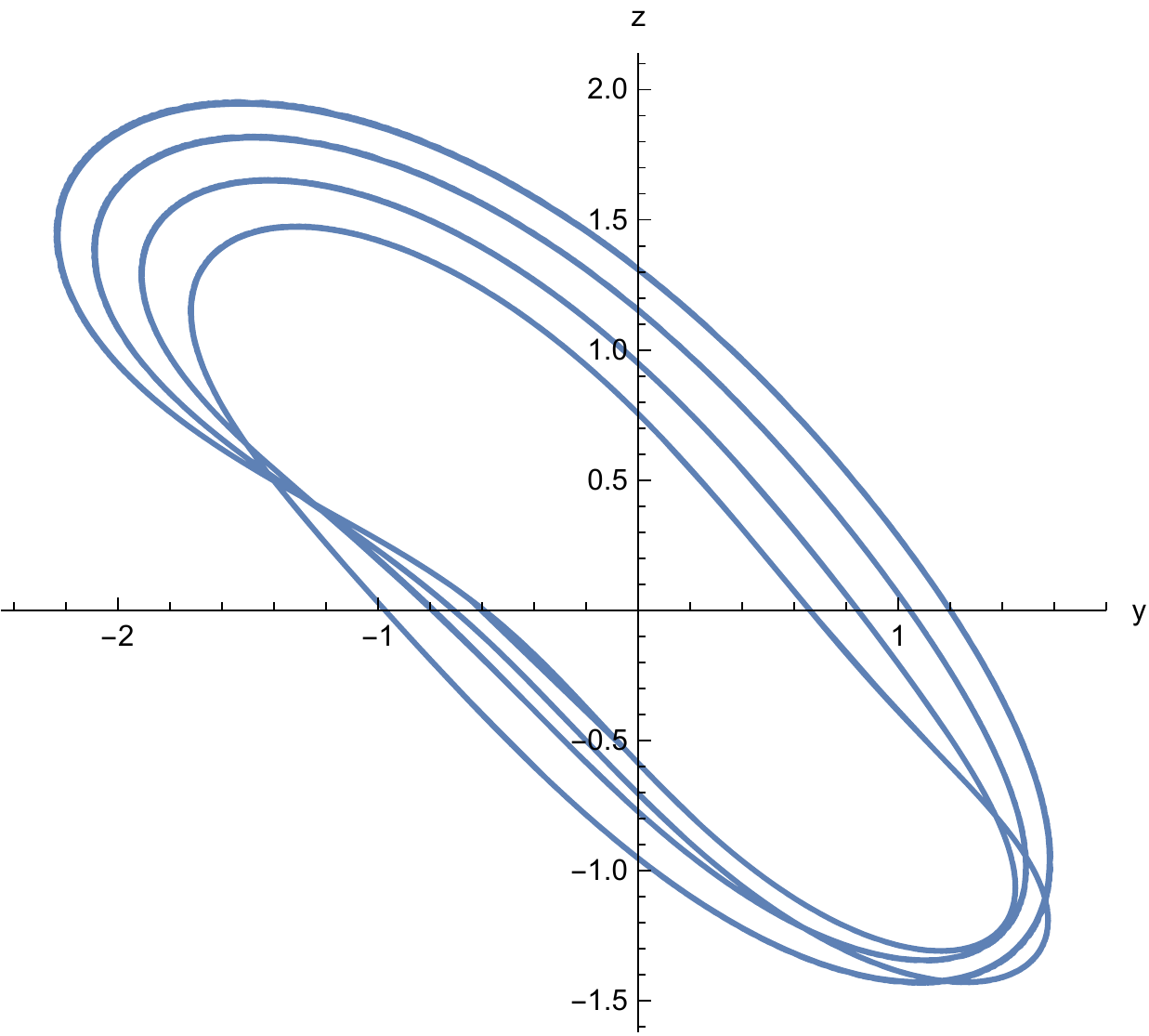}} 
		\\
		\subfloat[$b=5.5$]{\includegraphics[width=0.35\textwidth]{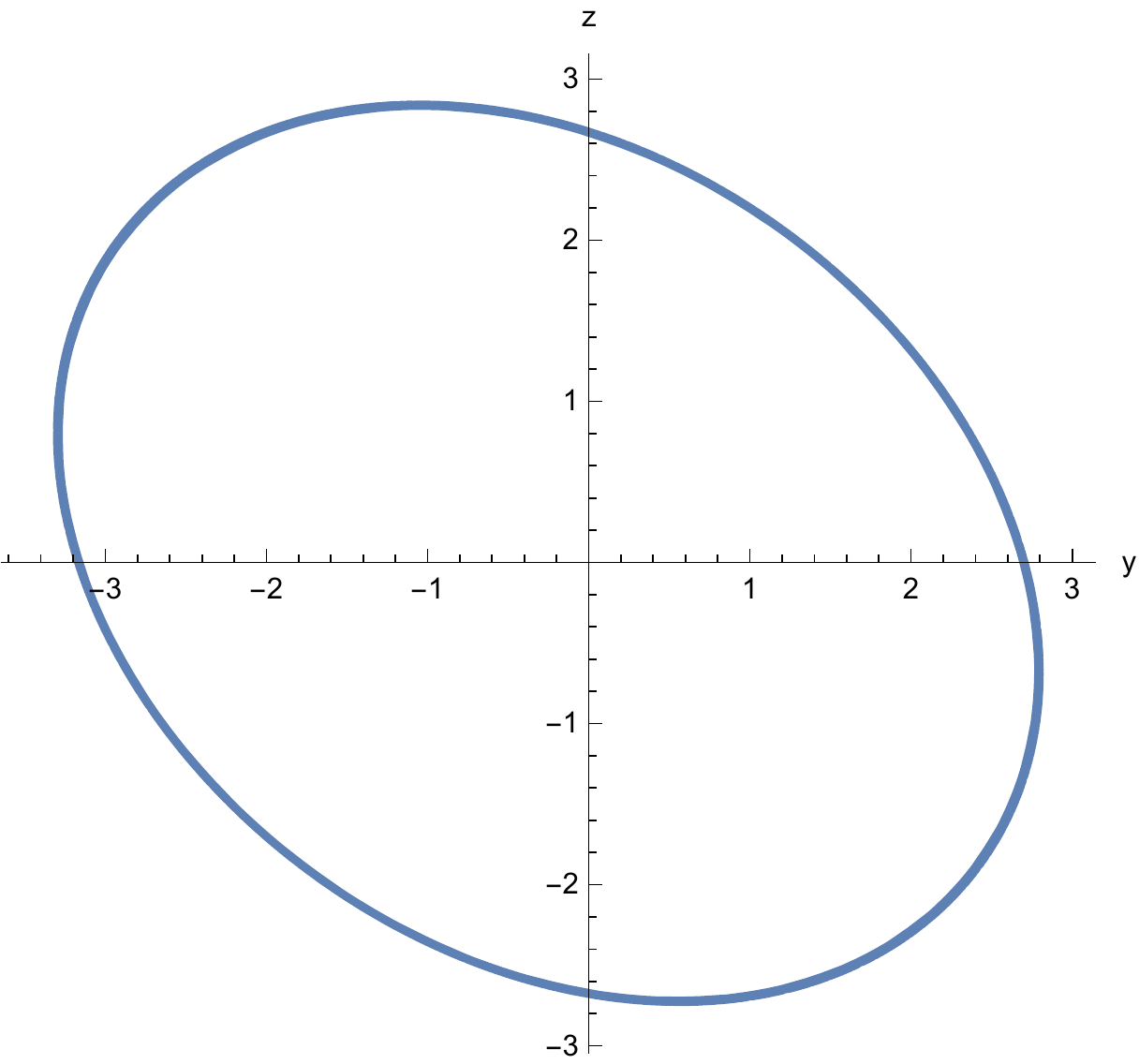}} 
		& 
		\subfloat[Bifurcation with parameter $b$]{\includegraphics[width=0.5\textwidth]{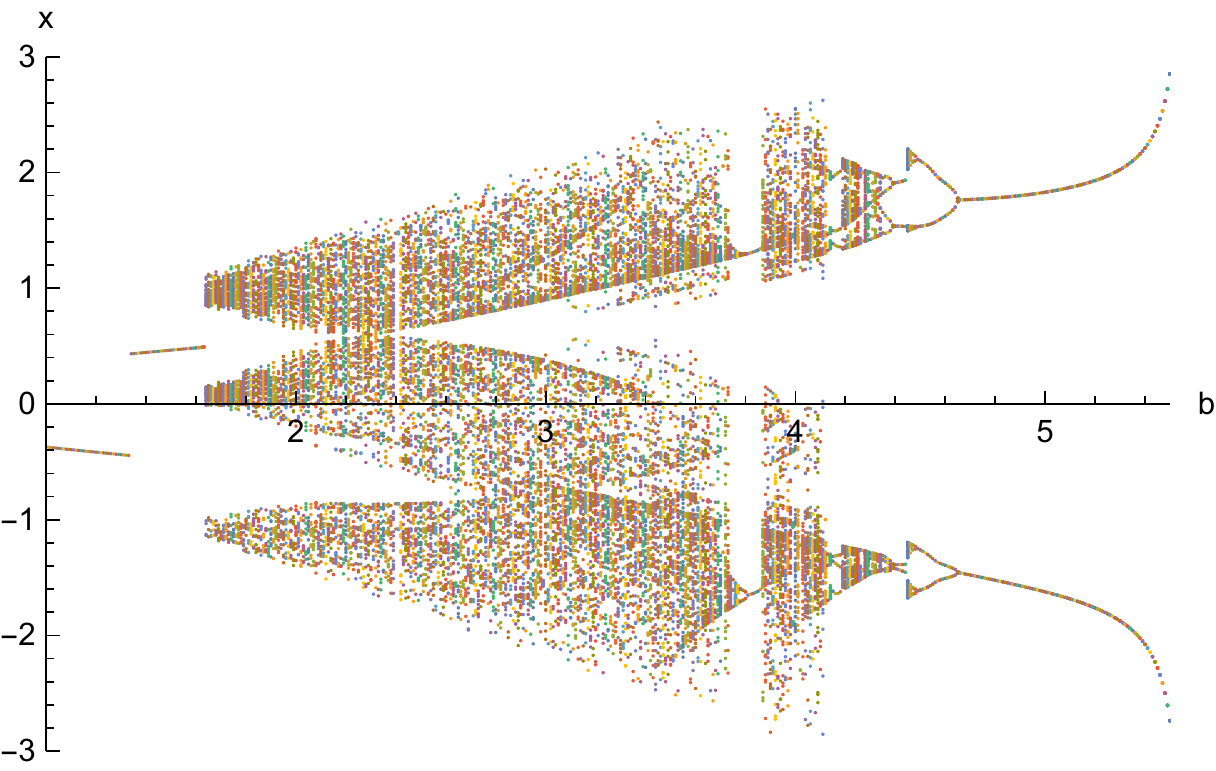}} 
	\end{tabular}
	\caption{Bifurcation analysis for parameter $b$}
	\label{Figb}
\end{figure*}

\begin{figure*}
	\begin{tabular}{c c }
		
		\subfloat[$g=2.7$  ]{\includegraphics[width=0.3\textwidth]{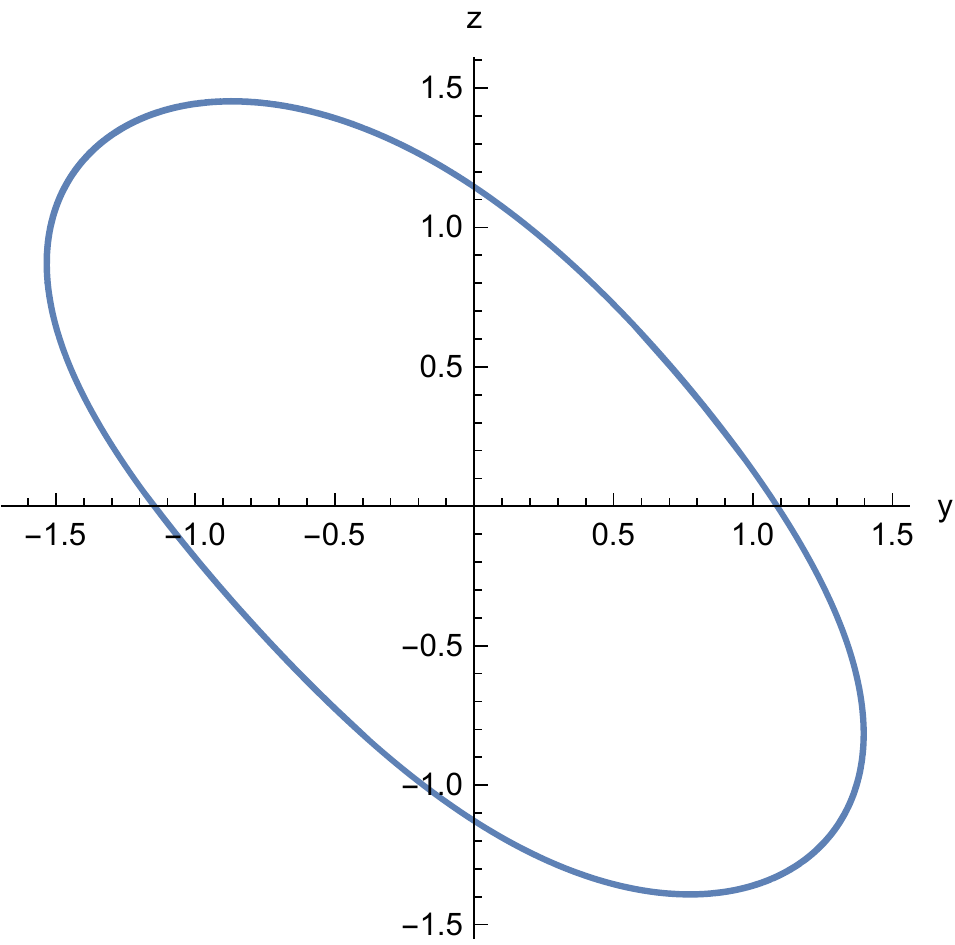}}
		&
		\subfloat[$g=3.4$ ]{\includegraphics[width=0.35\textwidth]{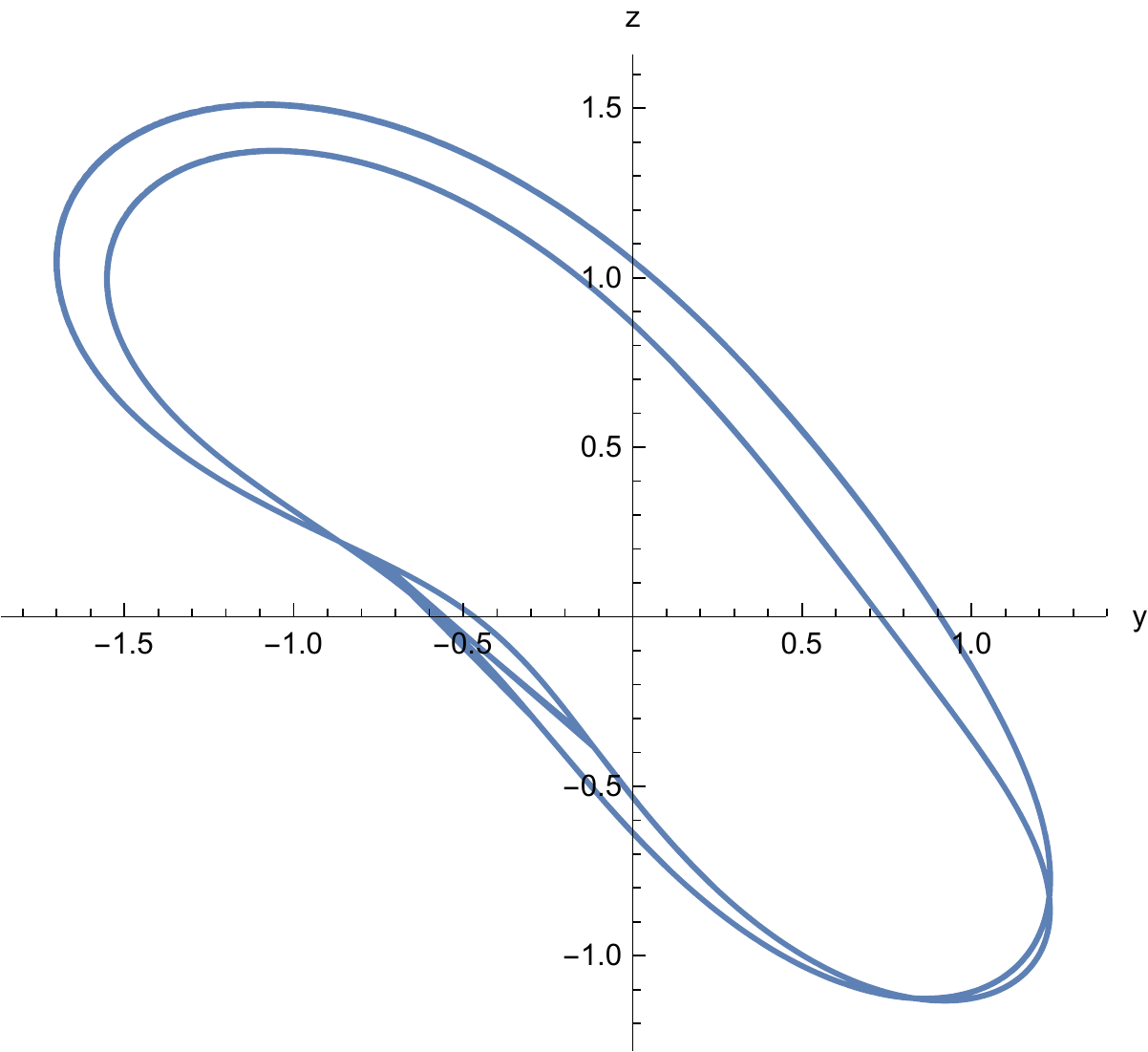}}	
		\\ 
		\subfloat[$g=3.8$]{\includegraphics[width=0.4\textwidth]{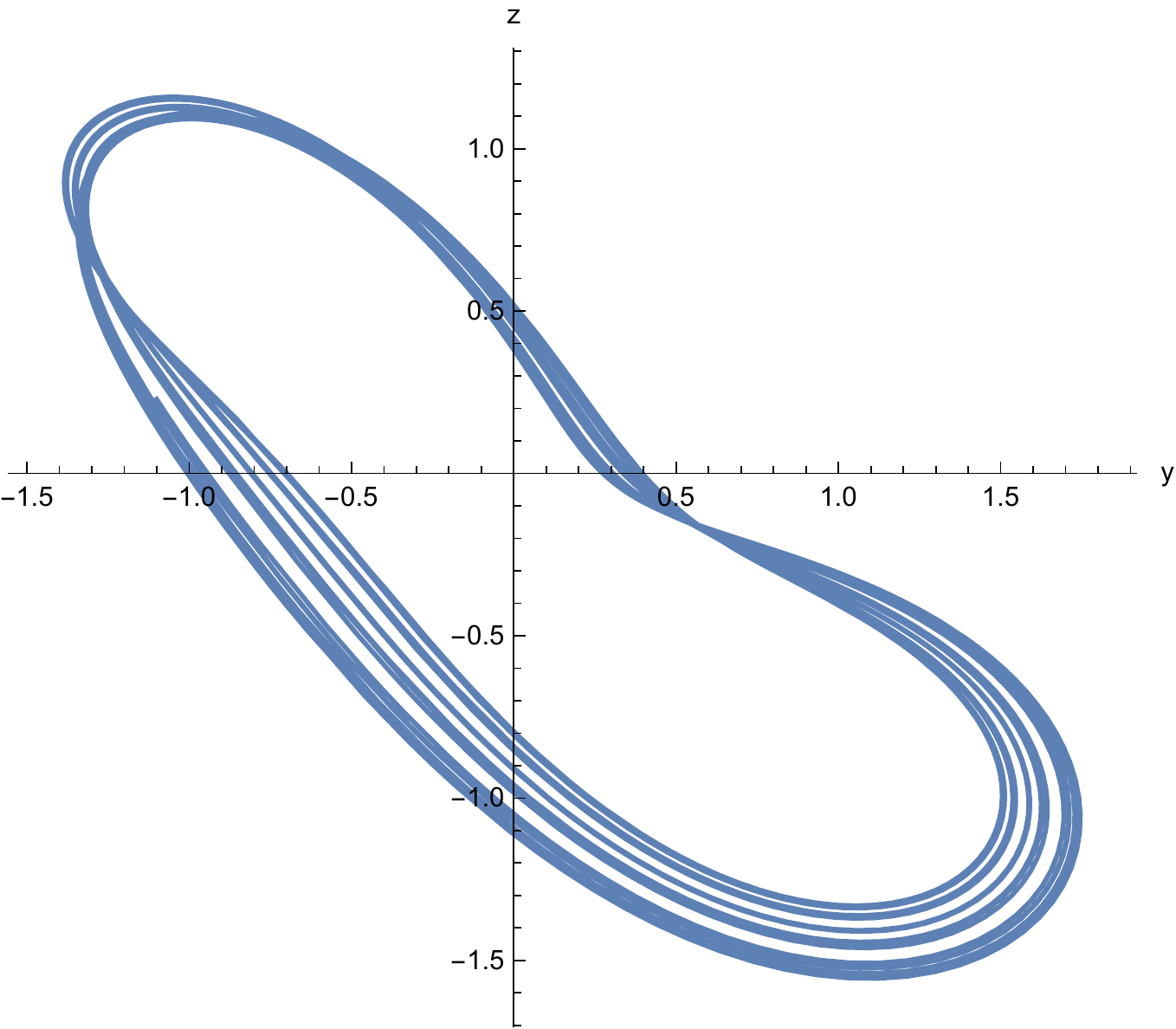}} 
		&
		\subfloat[$g=4.5$ ]{\includegraphics[width=0.15\textwidth]{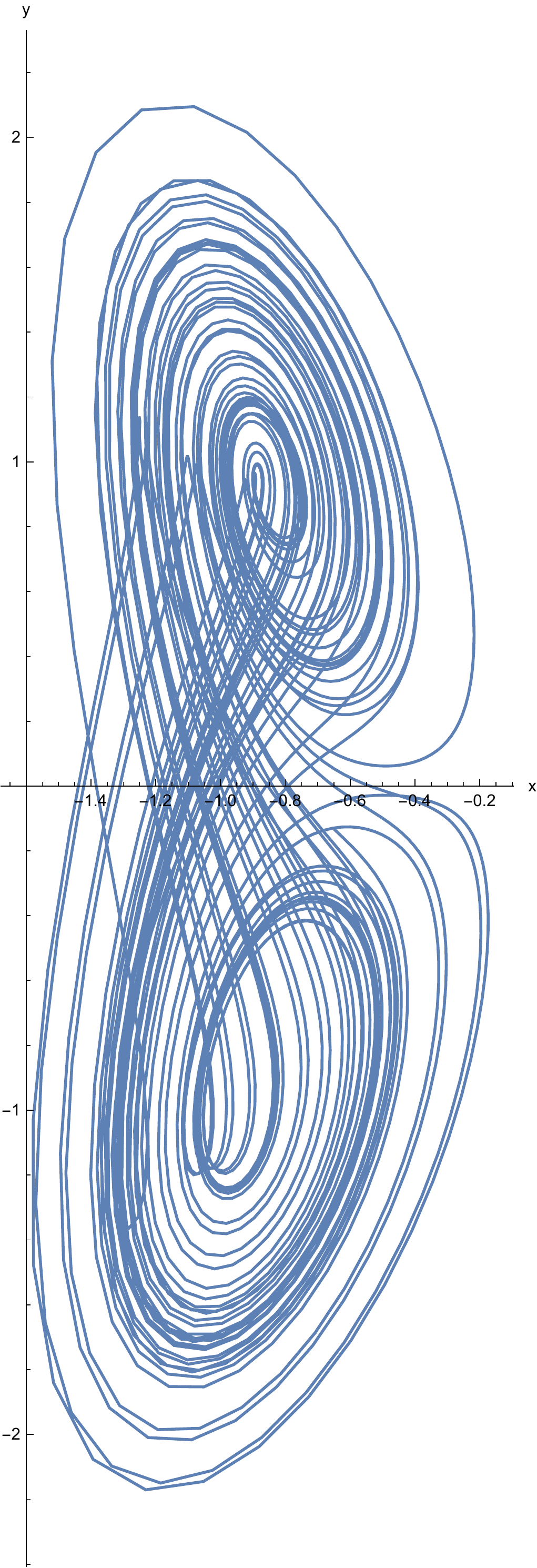}} 
		\\
		\subfloat[$g=5.5$]{\includegraphics[width=0.4\textwidth]{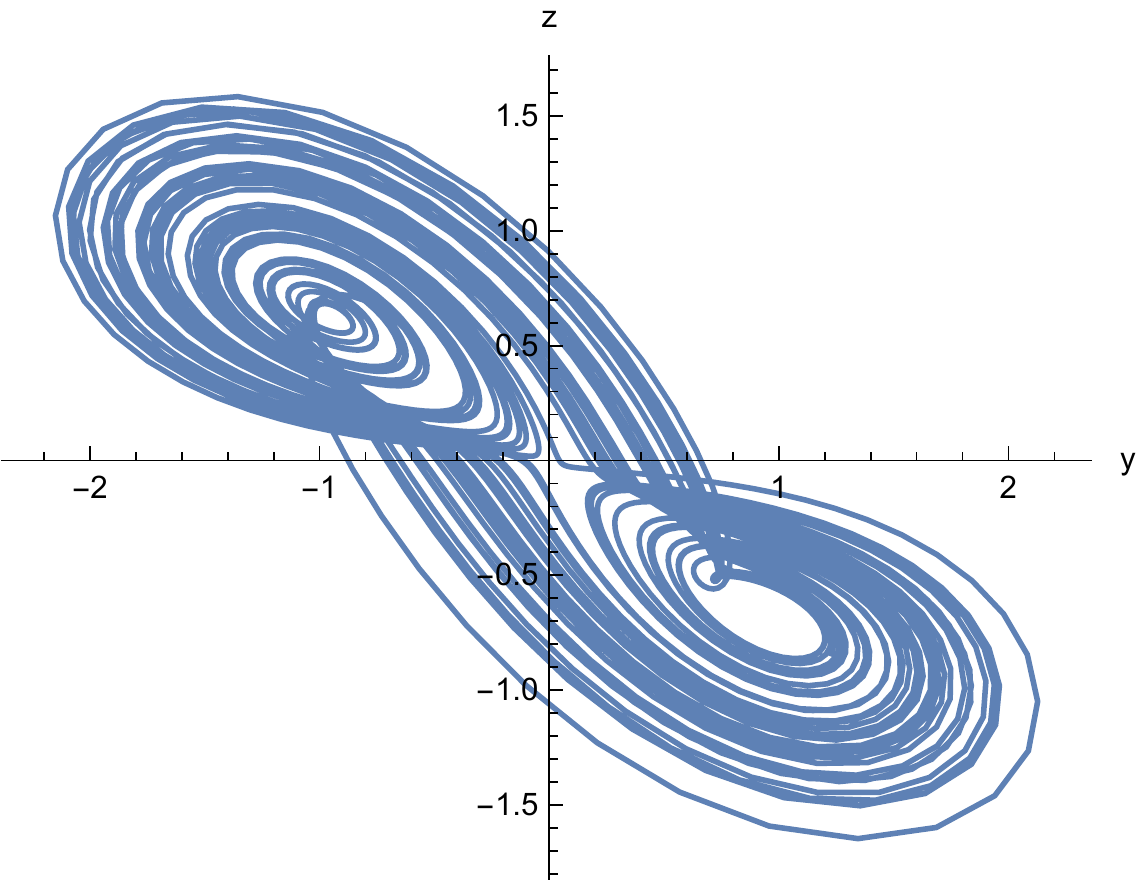}} 
		& 
		\subfloat[Bifurcation with parameter $g$]{\includegraphics[width=0.45\textwidth]{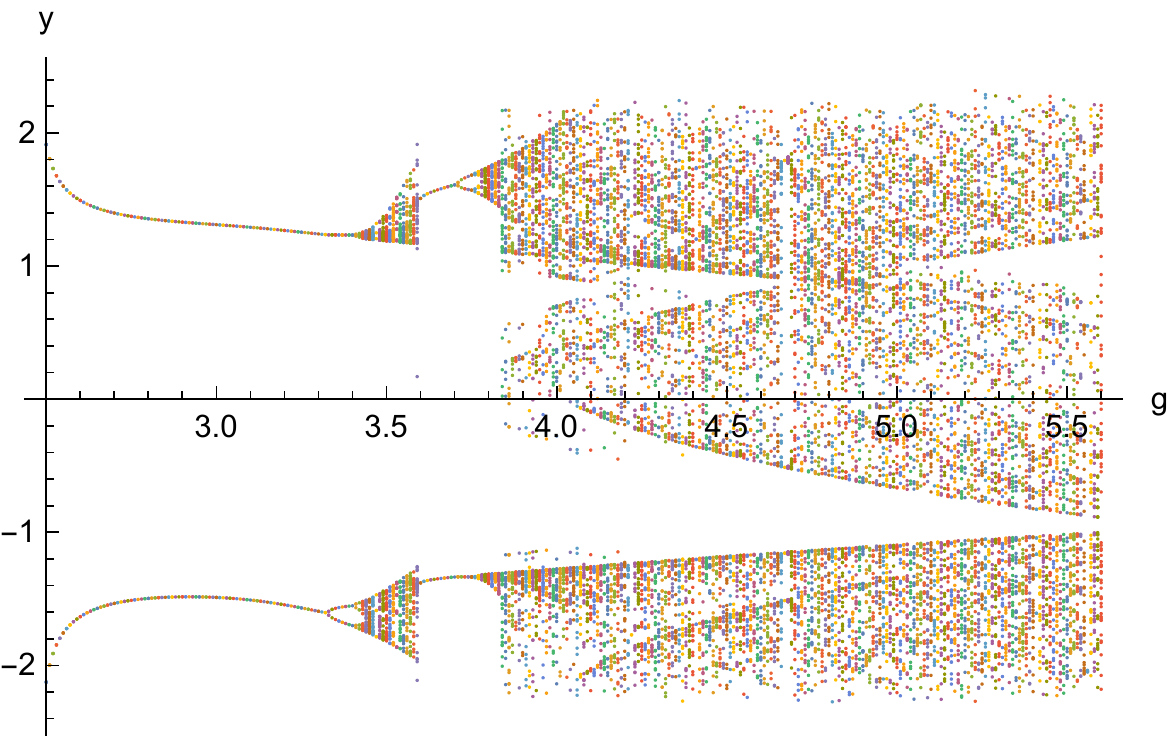}} 
	\end{tabular}
	\caption{Bifurcation analysis for parameter $g$}
	\label{Figg}
\end{figure*}

\begin{figure*}
	\begin{tabular}{c c }
		\subfloat[$h=-1.05$ ]{\includegraphics[width=0.55\textwidth]{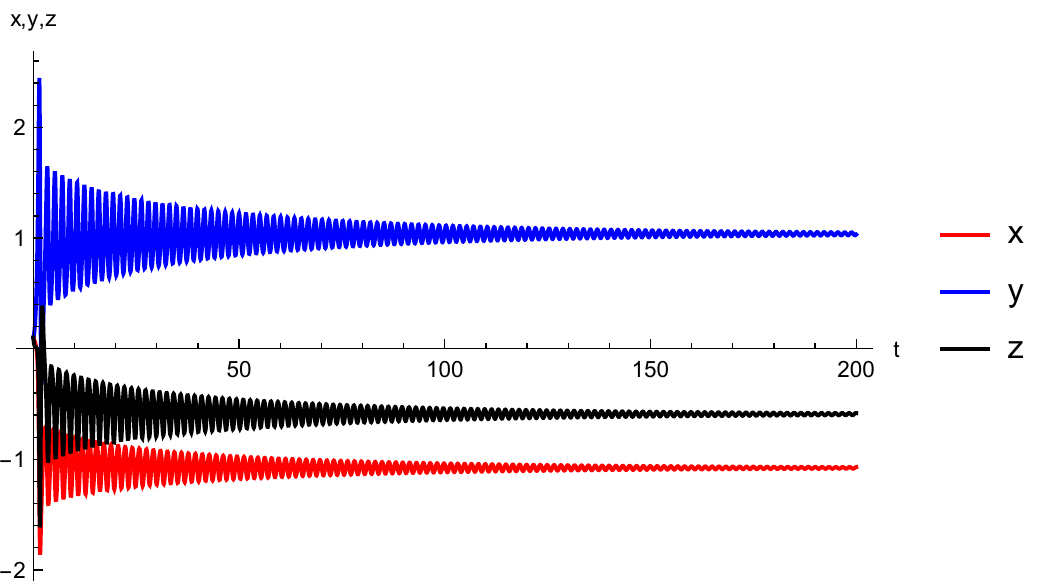}} 
		& 
		\subfloat[$h=-1$  ]{\includegraphics[width=0.25\textwidth]{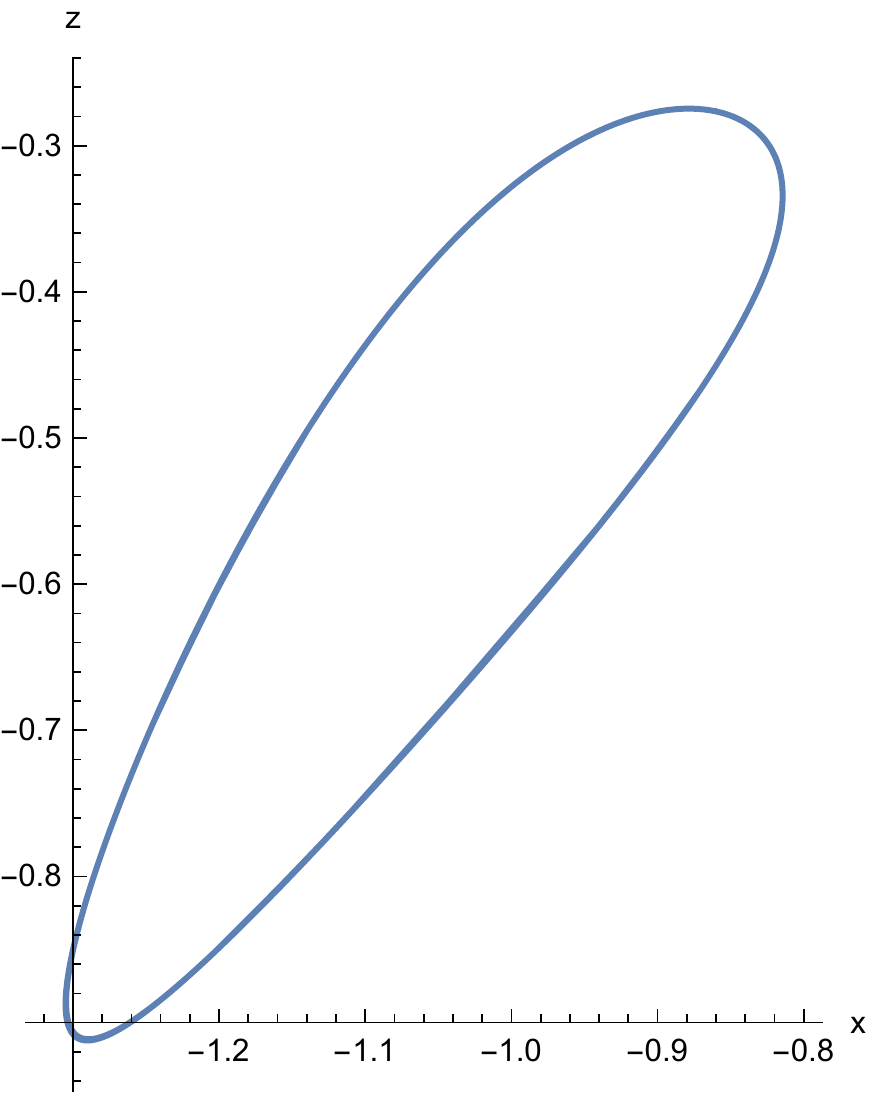}}
		\\
		\subfloat[$h=-0.2$ ]{\includegraphics[width=0.25\textwidth]{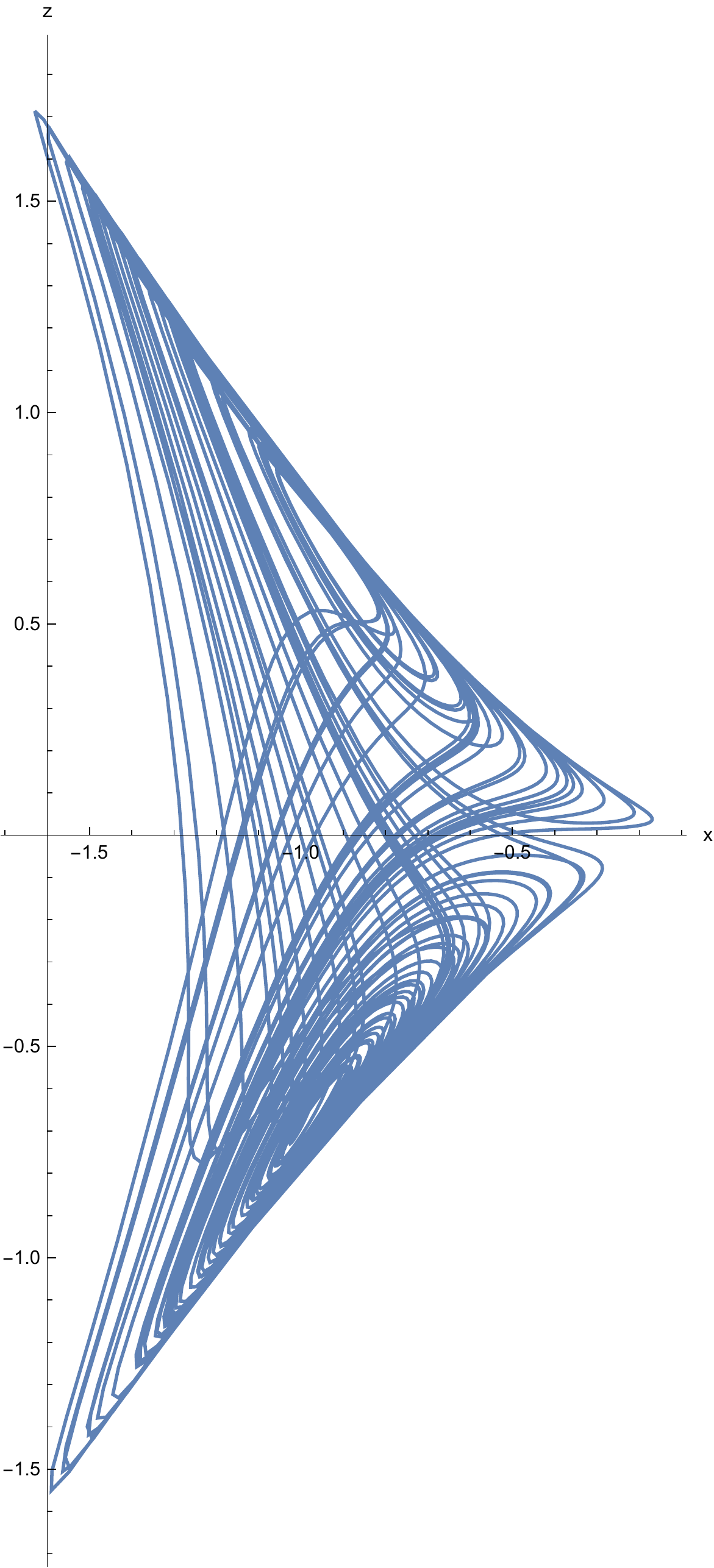}}	
		& 
		\subfloat[$h=0.5$]{\includegraphics[width=0.25\textwidth]{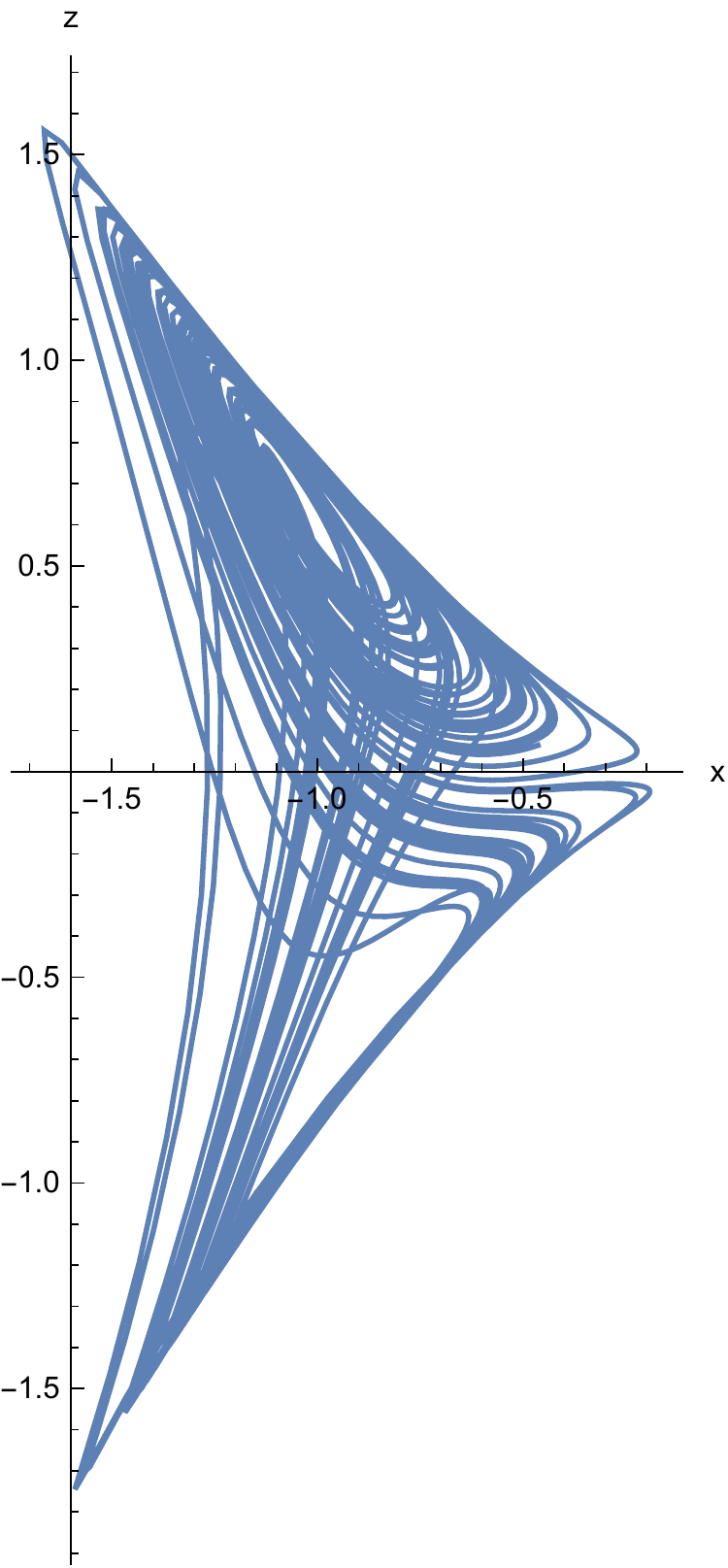}} 
		\\
		\subfloat[$h=0.97$]{\includegraphics[width=0.3\textwidth]{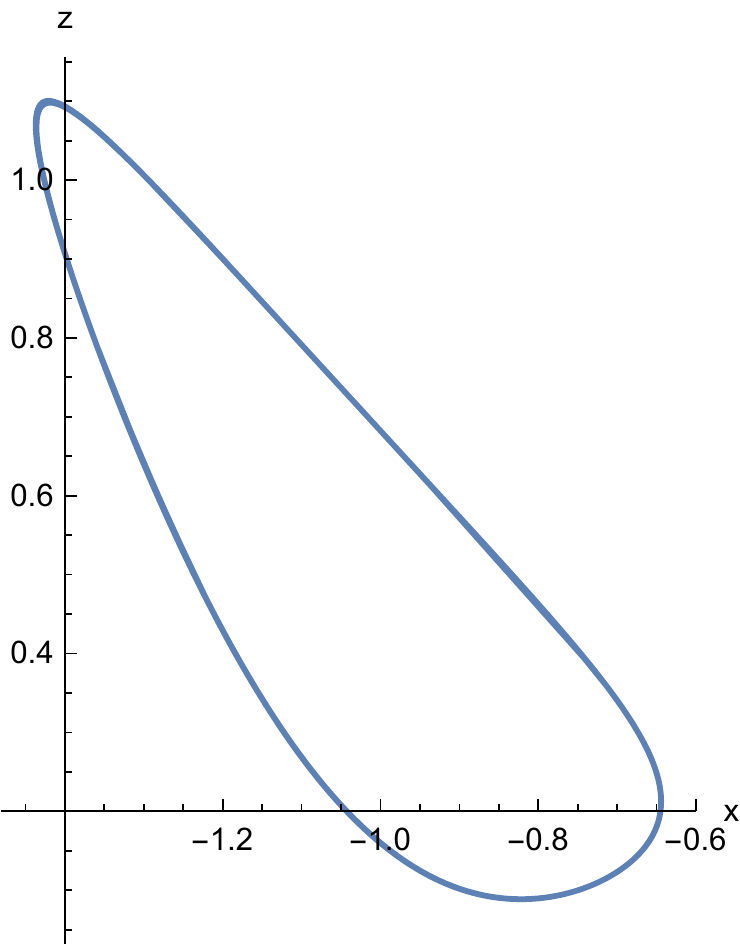}} 
		& 
		\subfloat[Bifurcation with parameter $h
		$]{\includegraphics[width=0.45\textwidth]{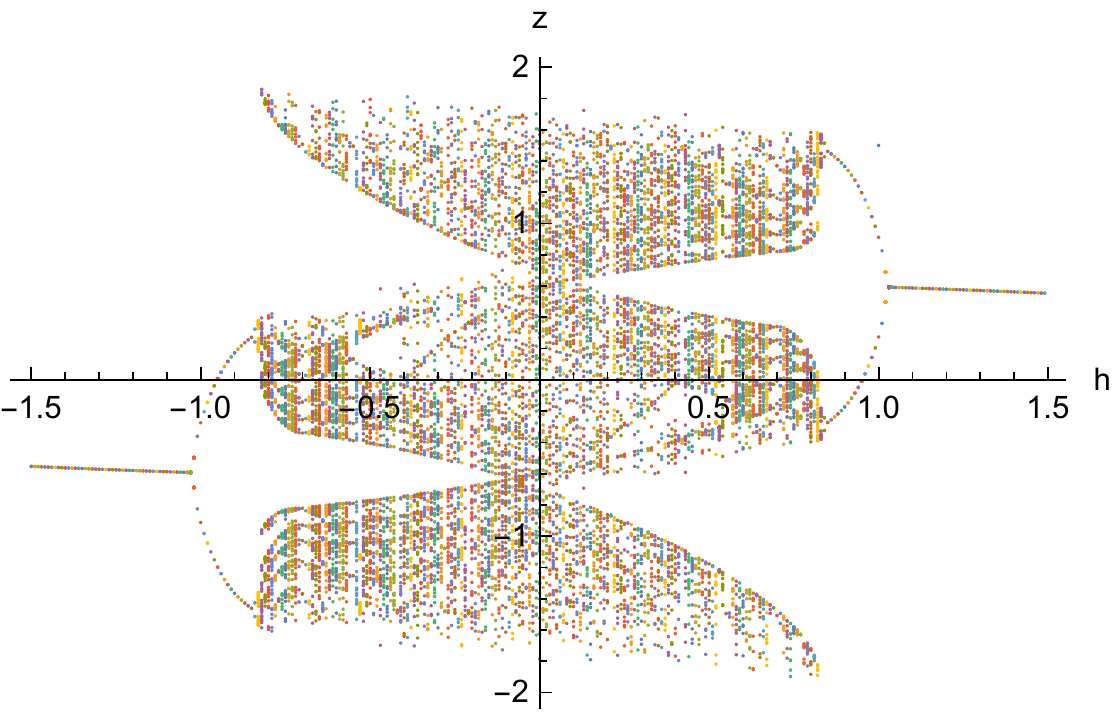}} 
	\end{tabular}
	\caption{Bifurcation analysis for parameter $h$}
	\label{Figh}
\end{figure*}

Out of five parameters $a$, $b$, $d$, $g$, $h$ of the system (\ref{5.1.1}), we hold any four parameters fixed and vary the remaining one to present the bifurcation analysis.    
 The bifurcation diagrams and the trajectories/phase portraits of the corresponding cases are presented in Figure \ref{Figa}- Figure \ref{Figh}. 

\begin{table}[h]
	\begin{center} 
		\begin{tabular}{|c|c|c|c|}
			\hline
		Fixed parameters & Changing parameter	& Behavior of trajectories & Corresponding figures \\
			\hline
$b=2.5$	&	$a=-4$ & Stable orbit & Figure \ref{Figa}(a) \\
			\cline{2-4}
	$d=-5$	&	$a=-3.5$ & Limit cycle & Figure \ref{Figa}(b)  \\
			\cline{2-4}
	$g=5.5$	&	$a=-2.8$ & Limit cycle & Figure \ref{Figa}(c)   \\
			\cline{2-4}
	$h=-0.2$	&	$a=-2.6$ & Limit cycle & Figure \ref{Figa}(d)  \\
			\cline{2-4}
		&	$a=-2$ & Chaos & Figure \ref{Figa}(e)  \\
			\hline
			\hline
	$a=-1$	&	$b=1.6$ & Stable orbit & Figure \ref{Figb}(a)  \\
			\cline{2-4}
	$d=-5$	&	$b=1.8$ & Chaos & Figure \ref{Figb}(b) \\
			\cline{2-4}
	$g=5.5$	&	$b=2.5$ & Chaos & Figure \ref{Figb}(c)   \\
			\cline{2-4}
	$h=-0.2$	&	$b=4.35$ & Limit cycle & Figure \ref{Figb}(d)  \\
			\cline{2-4}
		&	$b=5.5$ & Limit cycle & Figure \ref{Figb}(e) \\
			\hline
			\hline
	$b=2.5$	& $g=2.7$ & Limit cycle & Figure \ref{Figg}(b) \\
		\cline{2-4}
	$d=-5$	& $g=3.4$  &  Limit cycle & Figure \ref{Figg}(c)	\\
		\cline{2-4}
	$h=-0.2$	& $g=3.8$ & Limit cycle & Figure \ref{Figg}(d) \\
		\cline{2-4}
		$a=-1$	& $g=4.5$ & Chaos & Figure \ref{Figg}(a) \\
		\cline{2-4}
	& $g=5.5$ & Chaos & Figure \ref{Figg}(e)\\
	\hline
	\hline
$a=-1$	& $h=-1.05$ & Stable orbit & Figure \ref{Figh}(a) \\
	\cline{2-4}
	$b=2.5$	& $h=-1$ & Limit cycle & Figure \ref{Figh}(b) \\
	\cline{2-4}
$d=-5$	& $h=-0.2$  &  Limit cycle & Figure \ref{Figh}(c) \\
	\cline{2-4}
$g=5.5$	& $h=0.5$ & Limit cycle & Figure \ref{Figh}(d) 	\\
	\cline{2-4}
	& $h=0.97$ & Chaos & Figure \ref{Figh}(e) \\
	\hline
	
		\end{tabular}
		\caption{Observations for different values of parameters}
		\label{Tab2}
	\end{center}
\end{table}

\subsection{Equilibrium points and their nature}
 The equilibrium point of three dimensional dynamical system is called a saddle point of index $1$ (respectively, index $2$) if it generates one (respectively, two) unstable eigenvalue(s).
 The scrolls of chaotic attractors are usually around the saddle points of index $2$, whereas the saddles of index $1$ connect these scrolls \cite{Thompson,Cafagna}.
 \par In Table \ref{Tab1}, we classify equilibrium points of the proposed system (\ref{5.1.1}) with parameter values $a=-1$, $b=2.5$, $d=-5$, $g=5.5$ and $h=-0.2$. It is observed that the chaotic attractor in this system is in the neighborhood of the equilibrium points $O$, $E_1$ and $E_2$.

\begin{table}[h]
	\begin{center} 
 \begin{tabular}{|c|c|c|c|c|}
\hline
   & Equilibrium points $X_*$ & Eigenvalues of $J(X_*)$ & Nature of $X_*$\\
\hline
$O$ & $(0,0,0)$ & $-5.5, 2.5,-1$ & saddle point of index $1$\\
\hline
$E_1$ & $(-0.95606,-0.977783,0.646631)$ & $-4.67311, 0.336553\pm 3.61409 i$ & saddle point of index $2$\\
\hline
$E_2$ & $(-0.91615,0.957157,-0.668266)$ & $-4.86932, 0.434662\pm 3.57945 i$ & saddle point of index $2$\\
\hline
$E_3$ & $(-399.997,-19.9999,0.0250126)$ & $-399.997, -19.9999, 0.0250126$ & saddle point of index $2$\\
\hline
 \end{tabular}
  \caption{Equilibrium points of (\ref{5.1.1}), eigenvalues of Jacobian at these points and the nature of equilibriums}
 \label{Tab1}
\end{center}
\end{table} 
 
 \subsection{Chaos}

Consider the chaotic system,
\begin{equation}
\begin{split}
\dot{x}&=-x-y^2\\
\dot{y}&=2.5y-z-5xz\\
\dot{z}&=-5.5z+4xy-0.2x^2.
\end{split}\label{5.1}
\end{equation}
\par
\begin{figure}[h]
	\begin{center}
		\includegraphics[width=0.7\textwidth]{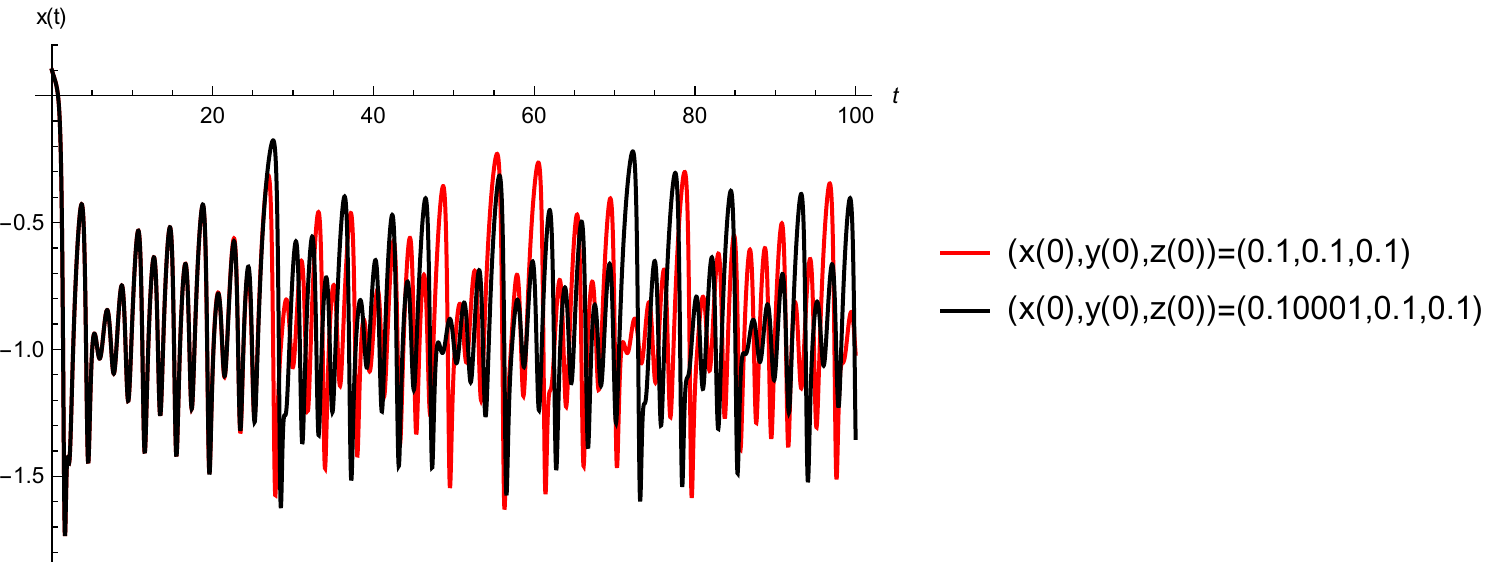}
		\caption{Solution trajectory $x(t)$ of the system (\ref{5.1}) with slightly different initial conditions.}         
		\label{Fig1}
	\end{center}  
\end{figure}

 We know that, extreme sensitivity to initial conditions is one of the significant characteristic of chaos. In the Figure \ref{Fig1}, we have plotted the trajectories $x(t)$ with initial conditions $(0.1,0.1,0.1)$ and $(0.10001,0.1,0.1)$ with small variation of $0.00001$. We can see that, there is huge amount of difference between these two trajectories. This high sensitivity to initial conditions indicates chaos in (\ref{5.1}).

\par
\begin{figure*}
	\begin{tabular}{c c}
		\subfloat[$z(t)$ ]{\includegraphics[width=0.5\textwidth]{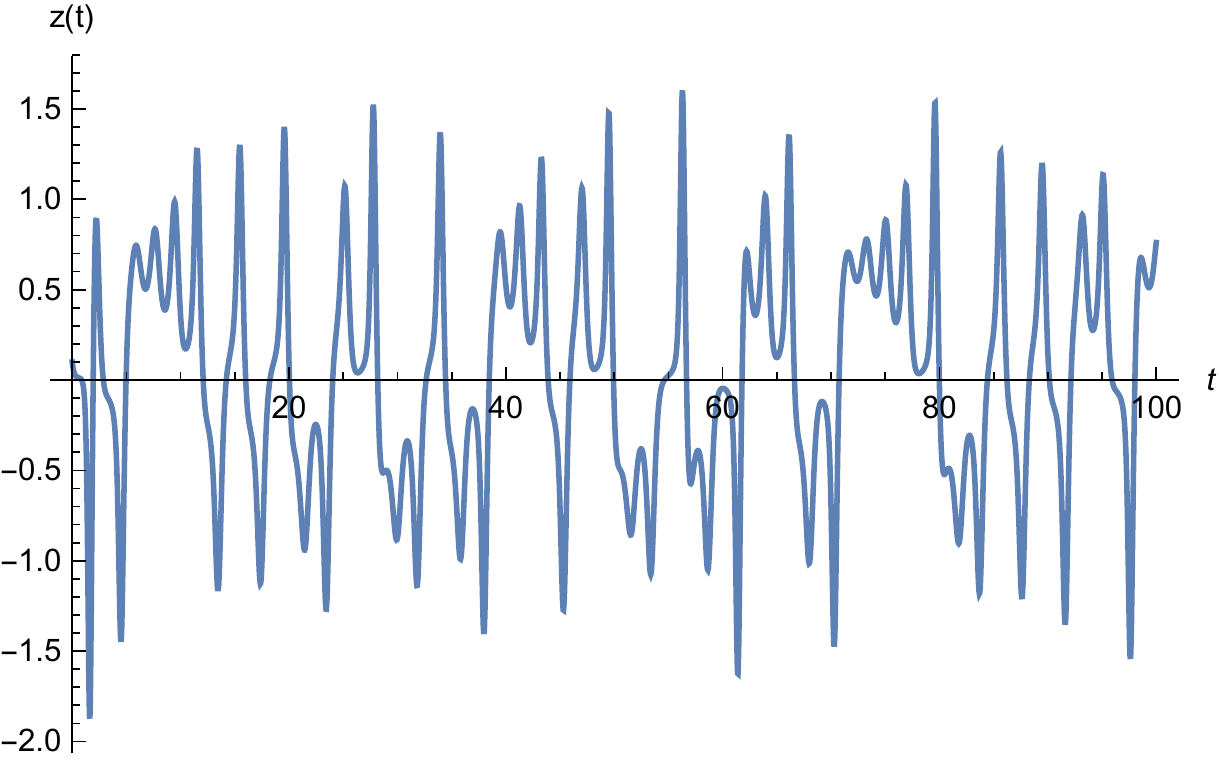}}
		& 
		\subfloat[$xy-plane$]{\includegraphics[width=0.2\textwidth]{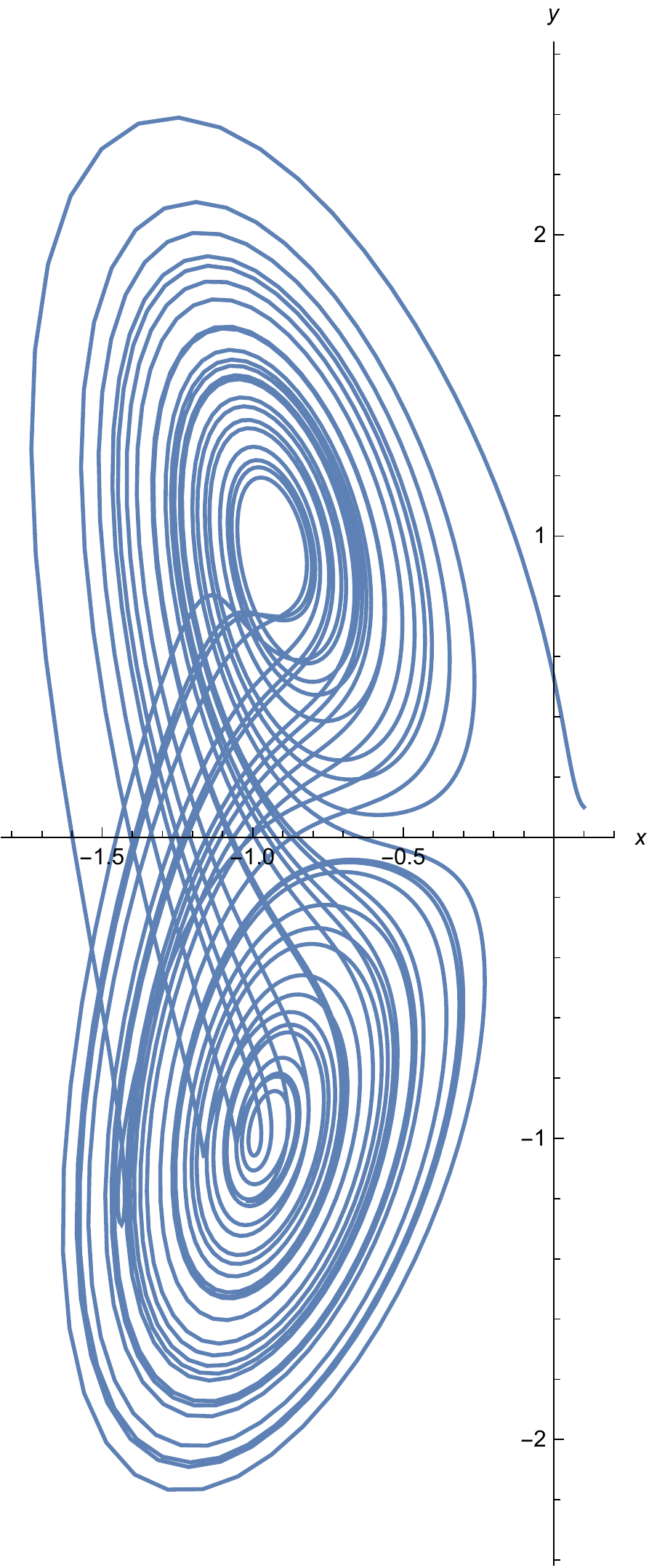}} 
	\end{tabular}
	\caption{Chaos in system (\ref{5.1})}
	\label{Fig2}
\end{figure*}

 Chaotic attractor and wave forms are shown in Figure \ref{Fig2}.

\subsection{Stability analysis of proposed fractional order system}

Consider the fractional order generalization of (\ref{5.1}) as
\begin{equation}
\begin{split}
{}_0^C\mathrm{D}_t^{\alpha}x &=-x-y^2\\
{}_0^C\mathrm{D}_t^{\beta}y &=2.5y-z-5x z\\
{}_0^C\mathrm{D}_t^{\gamma}z &=-5.5z+4x y-0.2 x^2.
\end{split} \label{5.7}
\end{equation}

Equilibrium points of fractional-order system (\ref{5.7}) and the eigenvalues of the Jacobian evaluated at these equilibrium points are same as their classical counterparts, described in Table \ref{Tab1}.
\subsubsection{Commensurate fractional order system}
Consider the fractional order system (\ref{5.7}) with $\alpha=\beta=\gamma$.
\par Define $$\alpha_*(E)=\frac{2}{\pi}\left[\min_{\substack{\lambda_E}}|\mathrm{arg(\lambda_E)}|\right]$$ as stability bound for the equilibrium point $E$. Therefore, $E$ is stable if $0<\alpha<\alpha_*(E)$ and unstable for $\alpha_*(E)<\alpha\leq 1$. Note that, instability of all the equilibrium points is the necessary condition \cite{Tavazoei-2007} to exists chaos in the commensurate order system (\ref{5.7}).
 \par In this case, 
\begin{equation*}
\begin{split}
\alpha_*(O)&=0,\\
\alpha_*(E_1)&=0.92307\\
 \mathrm{and} \,\, \alpha_*(E_2)&=0.940887.
\end{split}
\end{equation*}
We define 
\begin{equation}
\bar{\alpha}=\max_{\substack{E}} \alpha_*(E). \label{alb}
\end{equation}
In this case, $\bar{\alpha}=0.940887.$
If $\alpha<\bar{\alpha}$, then at least one of the equilibrium points $O$, $E_1$ or $E_2$ becomes stable and the system cannot be chaotic.
\par
\begin{figure*}
	\begin{tabular}{c c}
			
		\subfloat[$\alpha=0.97$, Chaotic trajectory ]{\includegraphics[width=0.6\textwidth]{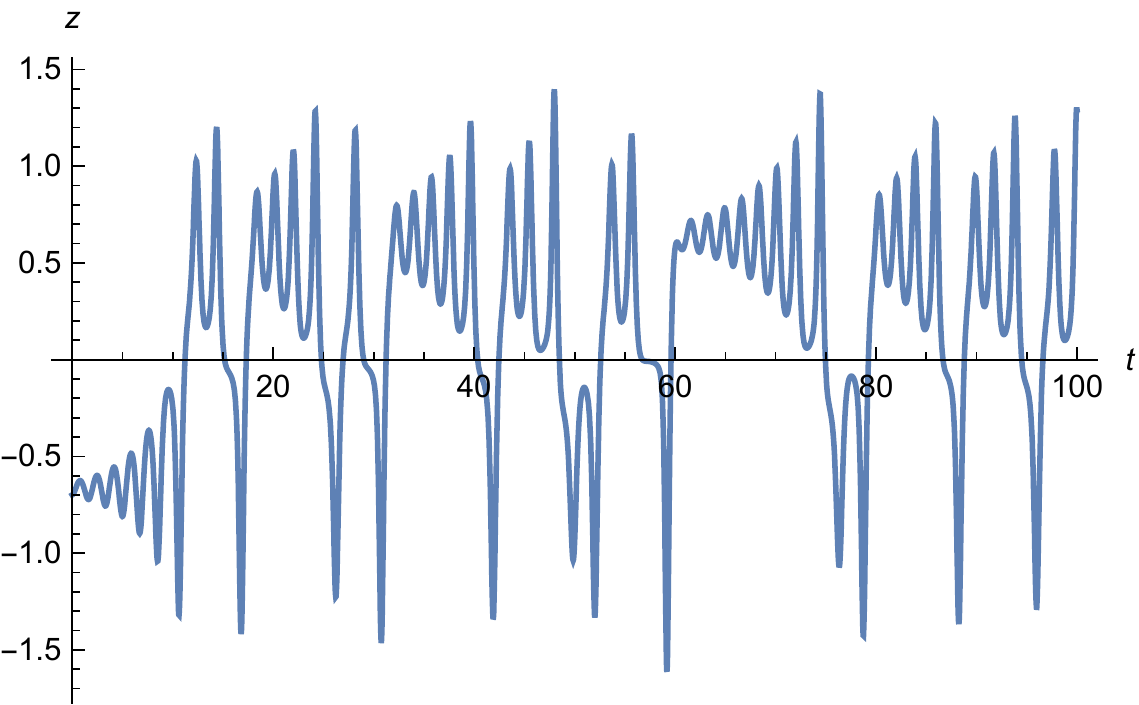}}
		&
		\subfloat[$\alpha=0.940887$, Chaotic attractor]{\includegraphics[width=0.2\textwidth]{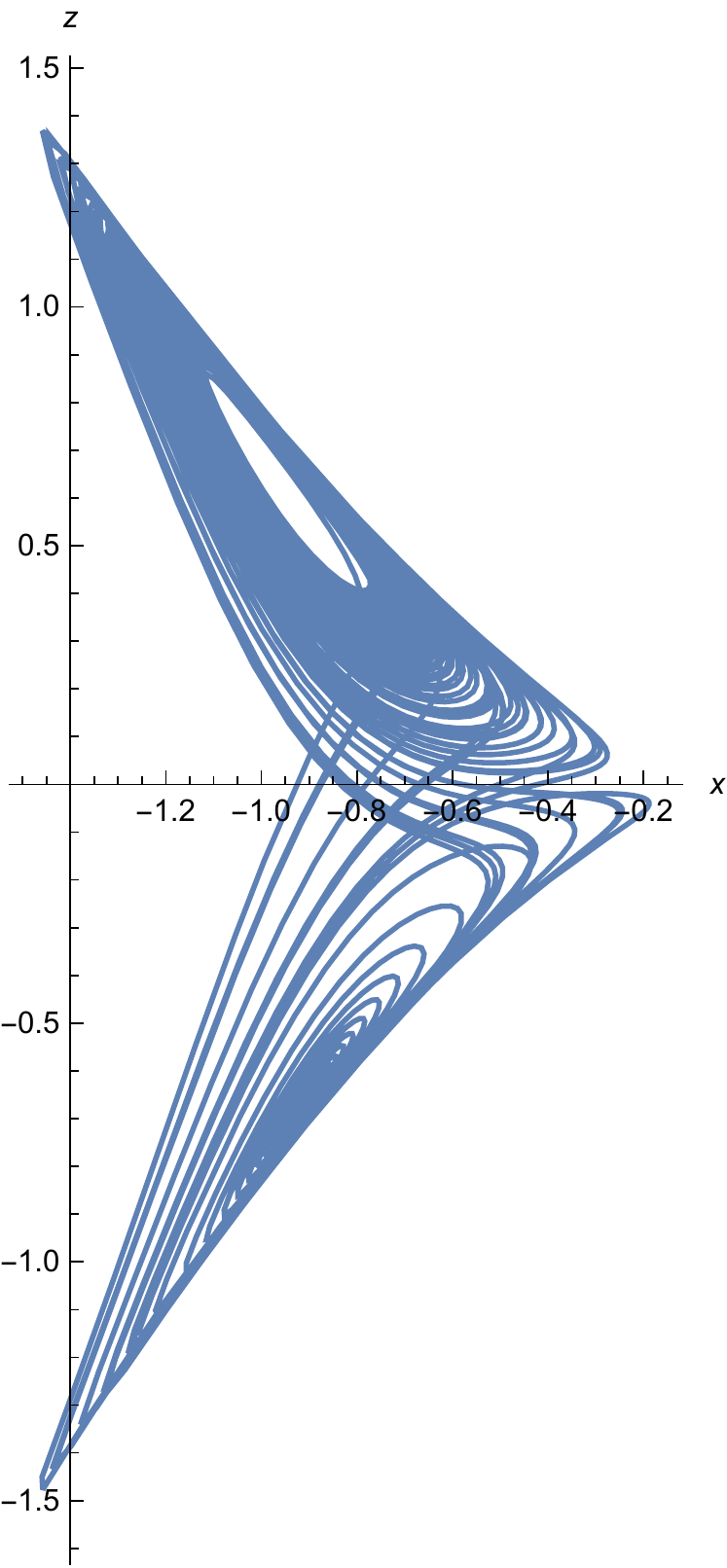}}  \\

		\subfloat[$\alpha=0.94$, stable ]{\includegraphics[width=0.6\textwidth]{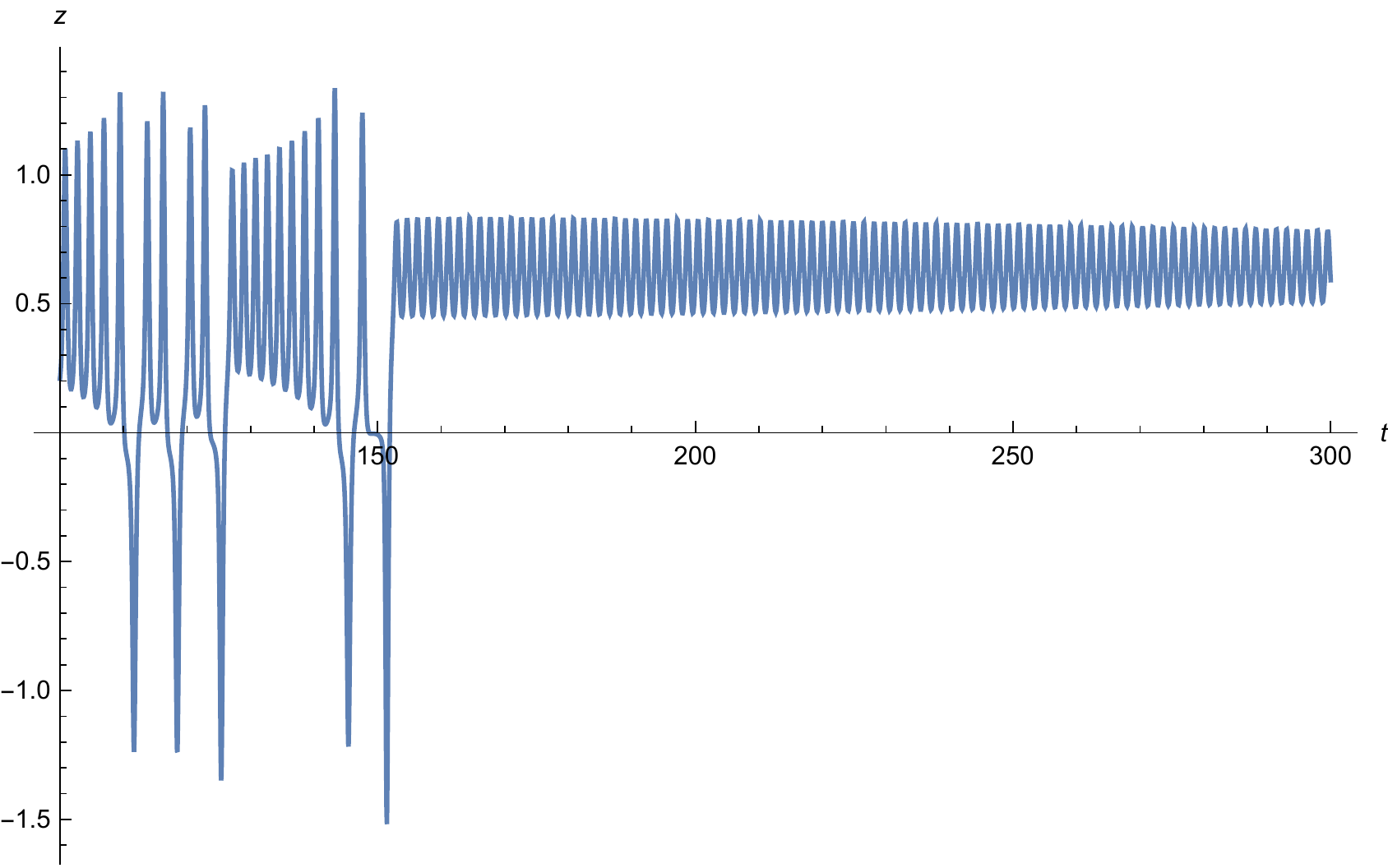}} 
		&
		\subfloat[$\alpha=0.93$, stable ]{\includegraphics[width=0.6\textwidth]{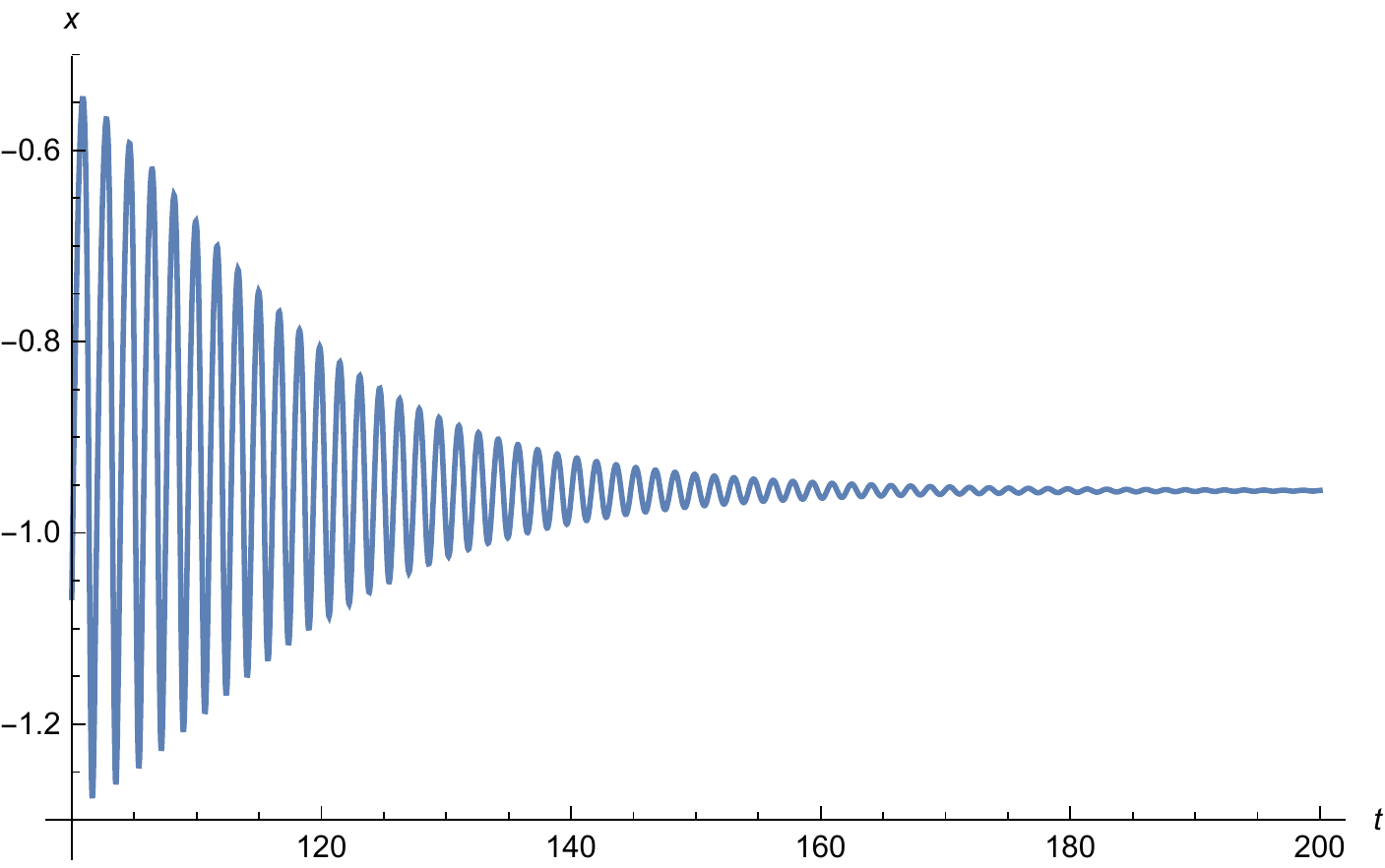}} 
	\end{tabular}
	\caption{Stability of commensurate fractional order system (\ref{5.7})}
	\label{Fig3}
\end{figure*}

 We further define the ``threshold value" $\alpha_t$ such that the system is chaotic for $\alpha>\alpha_t$ and achaotic for $\alpha<\alpha_t$. For the commensurate order system (\ref{5.7}) $\alpha_t=0.940887$ (cf. Figures \ref{Fig3} (a-d)).

\par Note that, $\alpha_t$ is the value obtained using numerical observations and $\alpha_t\geq \bar{\alpha}$.



Based on the observations of various fractional order commensurate order chaotic systems we propose the following conjecture.

\begin{conj}
Suppose that the system 
\begin{equation}
{}_0^C\mathrm{D}_t^{\alpha}X=f(X) \label{5.8}
\end{equation} 
is chaotic with the threshold value $\alpha_t$. If $\bar{\alpha}$ defined in (\ref{alb}) corresponds to the equilibrium points associated with the chaotic attractor then $\alpha_t=\bar{\alpha}$.
	\end{conj}

\subsubsection{Incommensurate fractional order system}
\begin{figure*}
	\begin{tabular}{c c}
		\subfloat[$\alpha=0.86$, $\beta=\gamma=1$ ]{\includegraphics[width=0.6\textwidth]{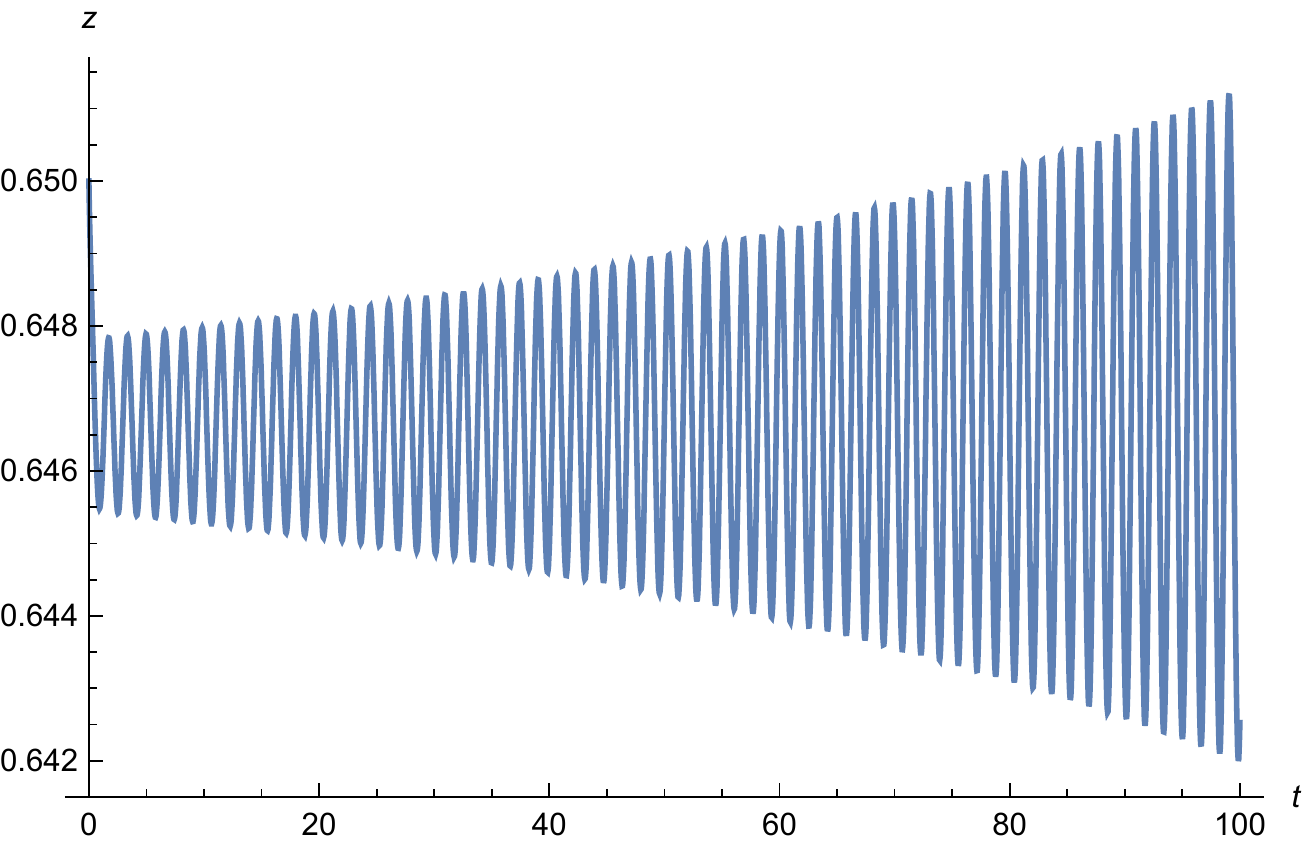}}
		& 
		\subfloat[$\alpha=0.9$, $\beta=\gamma=1$]{\includegraphics[width=0.35\textwidth]{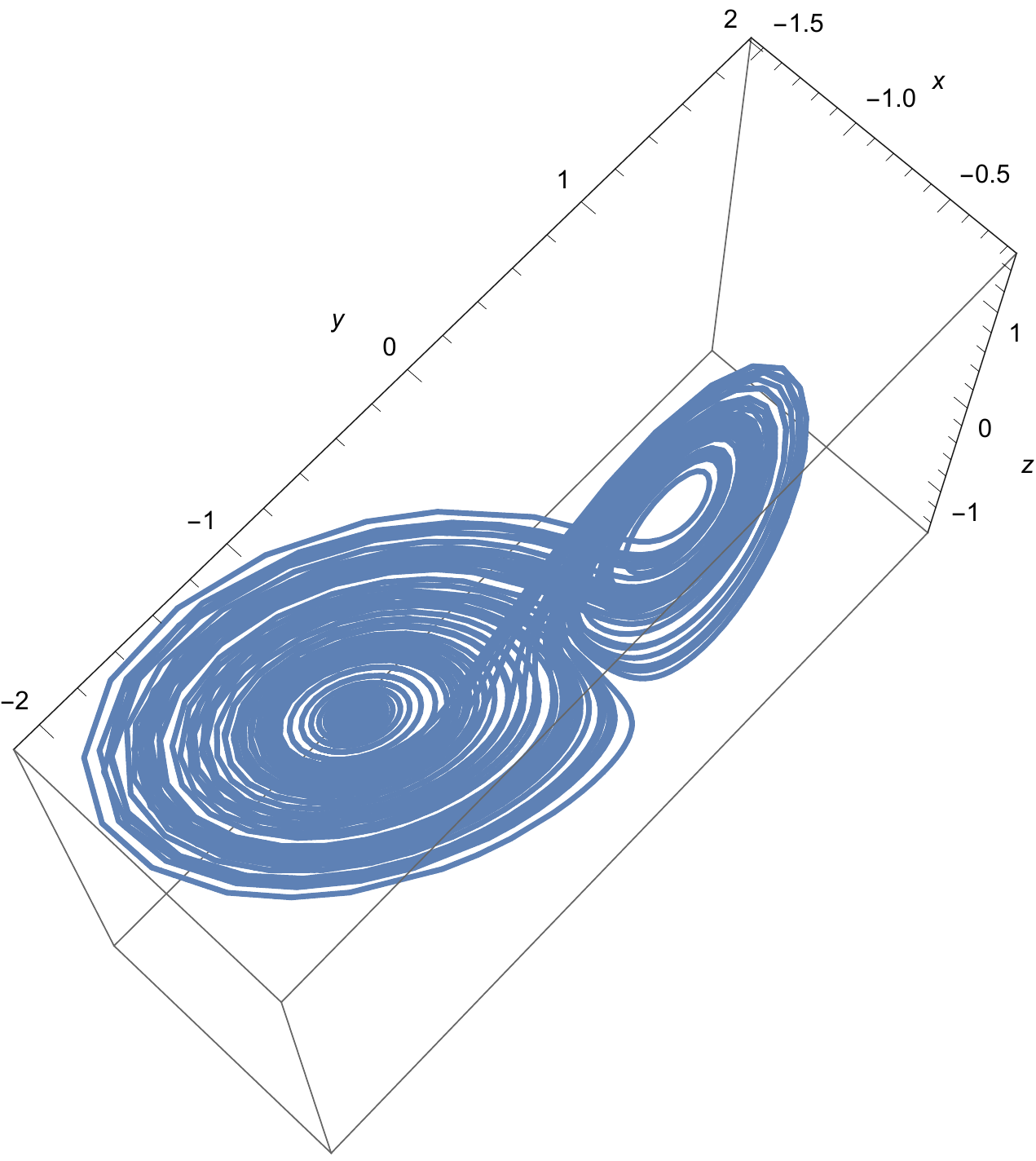}} \\
		
		\subfloat[$\gamma=0.85$, $\alpha=\beta=1$]{\includegraphics[width=0.52\textwidth]{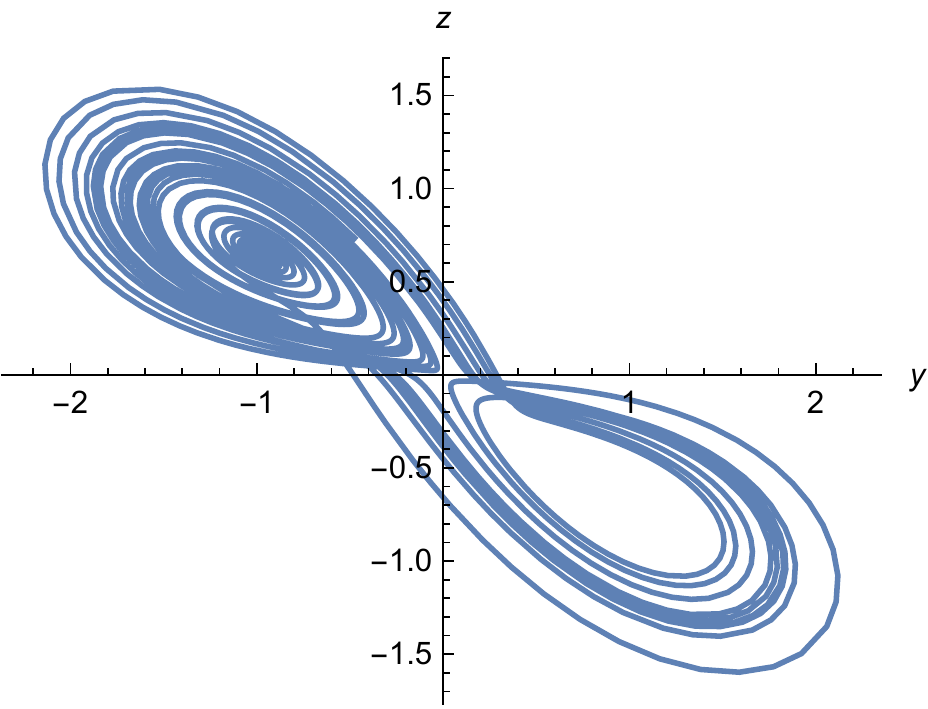}} 
		&	
	\end{tabular}
	\caption{Stability of incommensurate fractional order system}
	\label{Fig6}
\end{figure*}

We consider the system (\ref{5.7}) with two cases viz. $\alpha \ge 0.855$, $\beta=\gamma=1$ and $\alpha=\beta=1$, $\gamma \ge 0.68$.  We show that IMFOS $\ge 0 $ is not sufficient condition to exist chaos in the system. \\
{\bf Numerical simulations:}\\
{\bf Case 1:}\\
 Let $\alpha=\frac{43}{50}=0.86$ and $\beta=\gamma=1$.\\
 In this case $M=LCM(50,1,1)=50$, $\Delta_1=\mathrm{diag}(\lambda^{43},\lambda^{50},\lambda^{50})-J(O)$\\
$\mathrm{Det}(\Delta_1)=(1 + \lambda^{43}) (-2.5 + \lambda^{50}) (5.5 + \lambda^{50})$\\
IMFOS ($O$)  $= \frac{\pi}{100}-0 = 0.0314159 >0$\\
Similarly, we can find that,
IMFOS ($E_1$)  $= \frac{\pi}{100}-0.0313462 = 0.0000696784 >0$ and \\
IMFOS ($E_2$)  $= \frac{\pi}{100}-0.0308576 = 0.000558308 >0$\\
So, the system is unstable but there does not exist chaos (cf. Figure \ref{Fig6}(a)).\\

\noindent{\bf Case 2:}\\
Let $\alpha=\frac{9}{10}=0.9$ and $\beta=\gamma=1$.\\
In this case $M=LCM(10,1,1)=10$, $\Delta_2=\mathrm{diag}(\lambda^9,\lambda^{10},\lambda^{10})-J(O)$\\
$\mathrm{Det}(\Delta_2)=(1 + \lambda^9) (-2.5 + \lambda^{10}) (5.5 + \lambda^{10})$\\
IMFOS ($O$)  $= \frac{\pi}{20}-0 = 0.15708 >0$\\
Similarly, we can find that,
IMFOS ($E_1$)  $= \frac{\pi}{20}-0.154105 = 0.00297418 >0$ and \\
IMFOS ($E_2$)  $= \frac{\pi}{20}-0.151553 = 0.00552665 >0$\\
The system is chaotic (cf. Figure \ref{Fig6}(b)).\\

\noindent{\bf Case 3:}\\
Let $\gamma=\frac{85}{100}=\frac{17}{20}$ and $\alpha=\beta=1$.\\
In this case $M=LCM(20,1,1)=20$, $\Delta_3=\mathrm{diag}(\lambda^{20},\lambda^{20},\lambda^{17})-J(O)$\\
$\mathrm{Det}(\Delta_3)=(5.5 + \lambda^{17}) (-2.5 + \lambda^{20}) (1 + \lambda^{20})$\\
IMFOS ($O$)  $= \frac{\pi}{40}-0 = 0.0785398 >0$\\
Similarly, we can find that,
IMFOS ($E_1$)  $= \frac{\pi}{40}-0.0763924 = 0.00214747 >0$ and \\
IMFOS ($E_2$)  $= \frac{\pi}{40}-0.0748748 = 0.00366499 >0$\\
The system is chaotic (cf. Figure \ref{Fig6}(c)).\\

\section{Chaos control using linear control} \label{section5}
\begin{figure}[h]
	\begin{center}
		\includegraphics[width=0.7\textwidth]{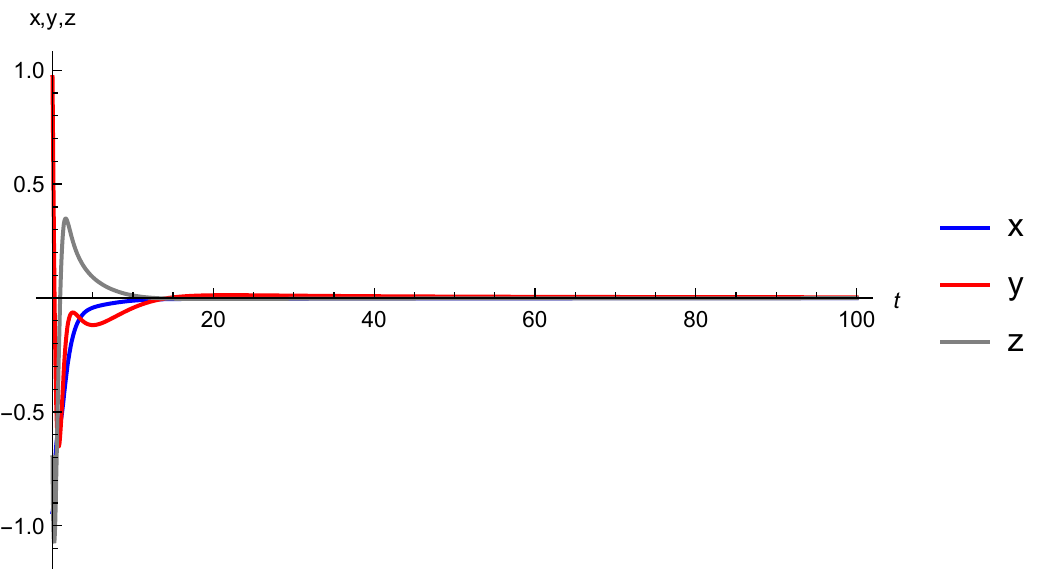}
		\caption{$\alpha=0.95$, $k=-3$, $X(0)=(-0.94, 0.97, -0.699)^T$}         
		\label{Fig9}
	\end{center}  
\end{figure}

Let us consider the controlled system,
\begin{equation}
\begin{split}
{}_0^C\mathrm{D}_t^{\alpha}x &=-x-y^2\\
{}_0^C\mathrm{D}_t^{\alpha}y &=2.5y-z-5x z+u\\
{}_0^C\mathrm{D}_t^{\alpha}z &=-5.5z+4x y-0.2 x^2,
\end{split}\label{5.9}
\end{equation}
where $u$ is the linear feedback control term. We set $u=ky$, where $k$ is a parameter to be determined so that the system (\ref{5.9}) is stable. Clearly $O=(0,0,0)$ is one of the equilibrium points of the system (\ref{5.9}). \\
The Jacobian of system (\ref{5.9}) evaluated at $O$ is, 
$$ \begin{bmatrix}
-1 &0 &0\\
0 & 2.5+k & -1\\
0 & 0 & -5.5
\end{bmatrix}. $$
Eigenvalues of this Jacobian matrix are $-1$, $2.5+k$ and $-5.5$. Therefore, if we choose $k < -2.5$ then the equilibrium point $O$ becomes stable and the system (\ref{5.9}) loose the chaotic nature. In  Figure \ref{Fig9} we observe stable behavior of system (\ref{5.9}) with $\alpha=0.95$ and $k=-3$.

\section{Chaos synchronization using optimal control} \label{section6}
We say that the chaotic systems get synchronized if the difference between their states tends to zero with increasing time. Synchronization between many pairs of fractional ordered chaotic systems are studied in the literature. Most of the fractional ordered chaotic systems are synchronized by using active control \cite{Bai,Bhalekar-control}, adaptive control \cite{Liao,Yassen}, sliding mode control \cite{Razminia,Muthukumar}, impulsive control \cite{Xing-Yuan,Jin-Gui}, projective control \cite{Wang,Peng}, etc. In this section, we propose a synchronization technique based on optimal control which depends on the fractional order of the system. 
\par
\begin{figure*}[h]
	\begin{tabular}{c c}
		\subfloat[Signals $x_1$, $x_2$ ]{\includegraphics[width=0.5\textwidth]{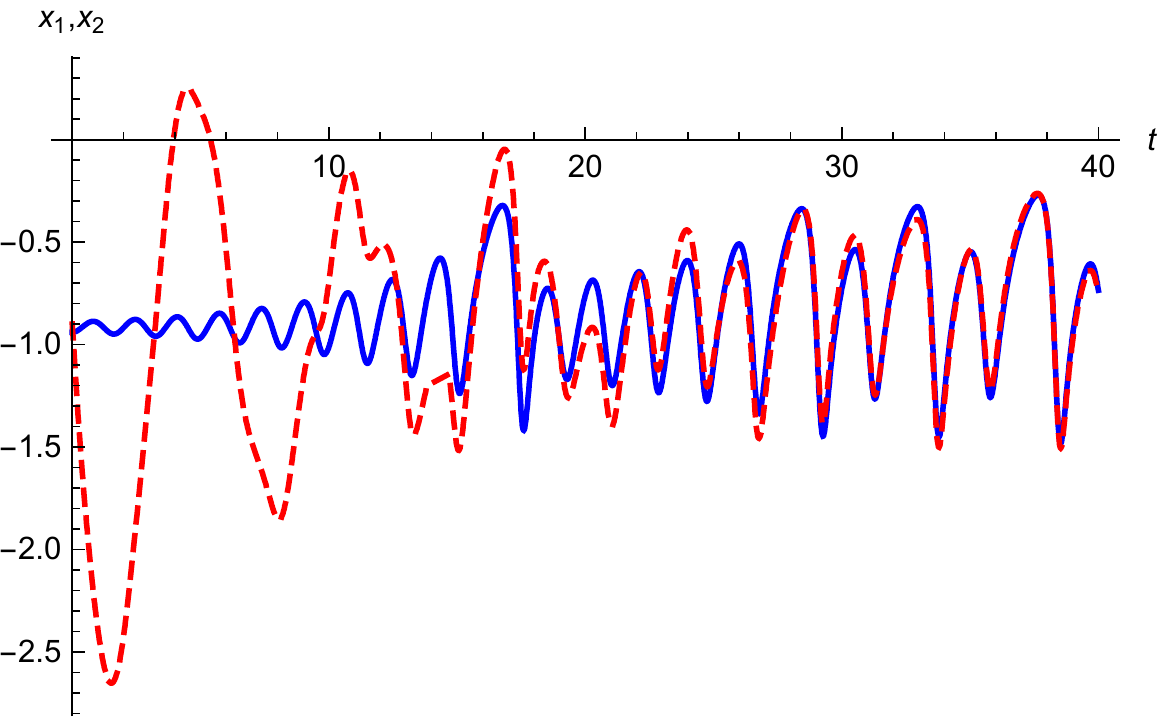}} 
		& 
		\subfloat[Signals $y_1$, $y_2$  ]{\includegraphics[width=0.5\textwidth]{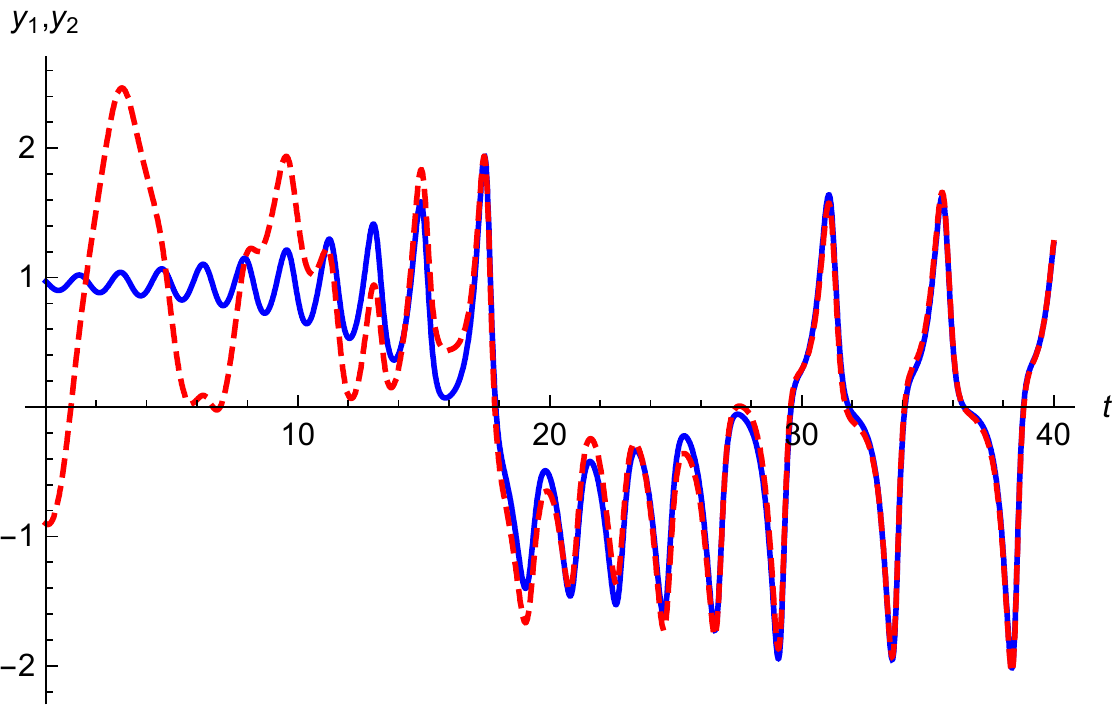}}
		\\
		\subfloat[Signals $z_1$, $z_2$ ]{\includegraphics[width=0.5\textwidth]{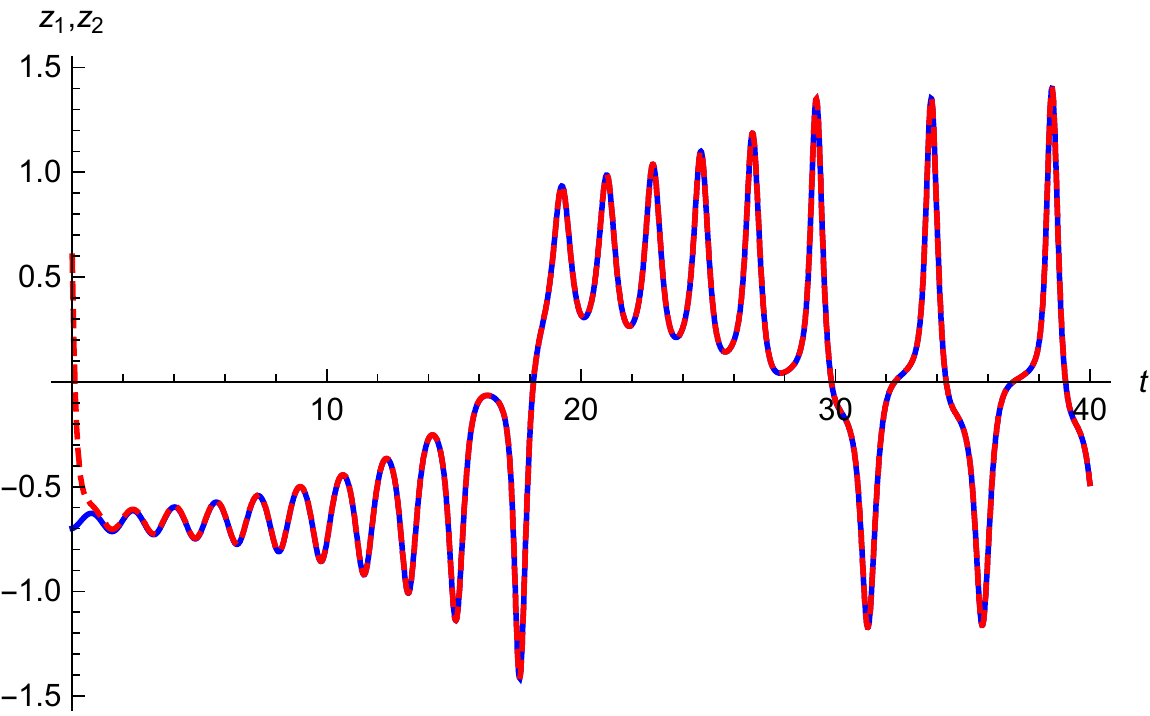}}
		&
		\subfloat[error system ]{\includegraphics[width=0.5\textwidth]{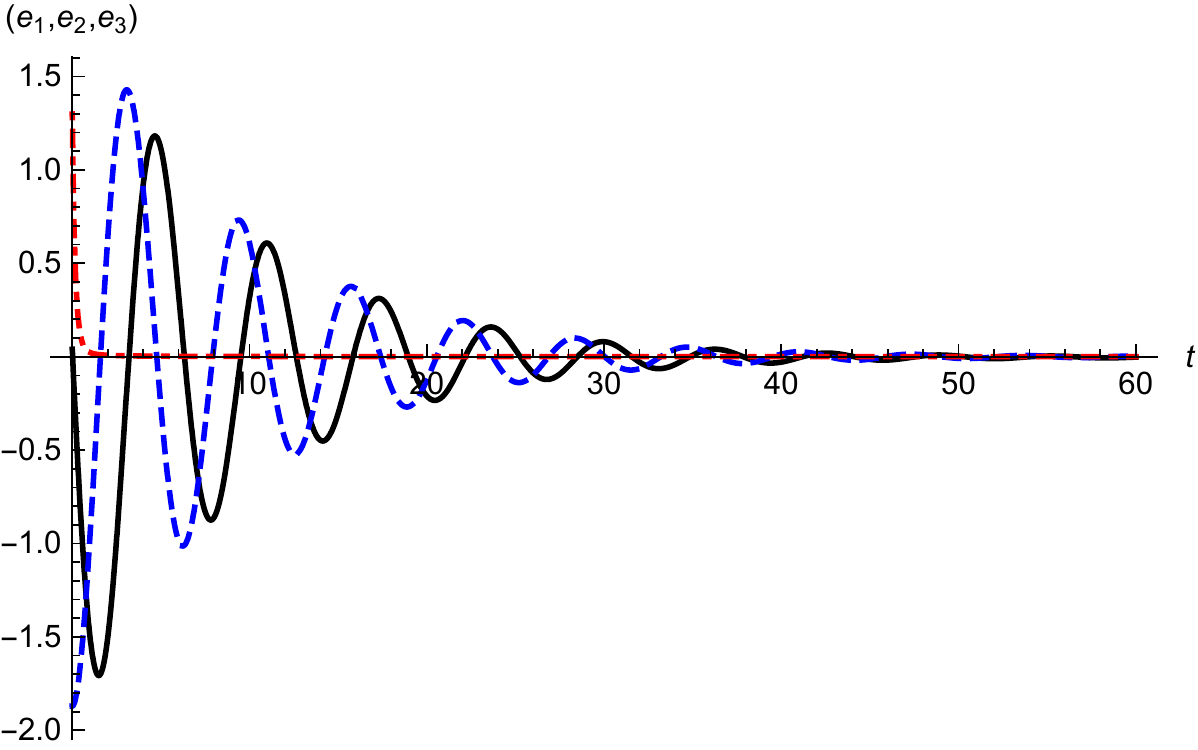}}
	\end{tabular}
	\caption{Synchronization of chaos in the proposed system}
	\label{Fig10}
\end{figure*}

 Here we have taken the system (\ref{5.7}) as both drive and response system.\\
Let us consider a drive system as 
\begin{equation}
\begin{split}
{}_0^C\mathrm{D}_t^{\alpha}x_1 &=-x_1-y_1^2\\
{}_0^C\mathrm{D}_t^{\alpha}y_1 &=2.5y_1-z_1-5x_1 z_1\\
{}_0^C\mathrm{D}_t^{\alpha}z_1 &=-5.5z_1+4x_1 y_1-0.2 x_1^2
\end{split} \label{5.10}
\end{equation}
 and response system as
\begin{equation}
\begin{split}
{}_0^C\mathrm{D}_t^{\alpha}x_2 &=-x_2-y_2^2+u_1\\
{}_0^C\mathrm{D}_t^{\alpha}y_2 &=2.5y_2-z_2-5x_2 z_2+u_2\\
{}_0^C\mathrm{D}_t^{\alpha}z_2 &=-5.5z_2+4x_2 y_2-0.2 x_2^2+u_3.
\end{split} \label{5.11}
\end{equation}
We define the error functions as,
\begin{equation}
e_1=x_1-x_2, \quad e_2=y_1-y_2, 
\quad e_3=z_1-z_2 \label{5.12}
\end{equation}
The error system can be given by using equations (\ref{5.10}), (\ref{5.11}) and (\ref{5.12}) as
\begin{equation}
\begin{split}
{}_0^C\mathrm{D}_t^{\alpha}e_1 &=-e_1+y_2^2-y_1^2-u_1\\
{}_0^C\mathrm{D}_t^{\alpha}e_2 &=2.5e_2-e_3-5x_1z_1+5x_2 z_2-u_2\\
{}_0^C\mathrm{D}_t^{\alpha}e_3 &=-5.5e_3+4x_1 y_1-4x_2y-2+0.2 x_2^2-0.2x_1^2-u_3.
\end{split} \label{5.13}
\end{equation}
Let us choose the control terms $u_i(t)$ in system (\ref{5.11}) as,
\begin{equation}
\begin{split}
u_1 &=-e_1+y_2^2-y_1^2-\cos(\frac{\alpha \pi}{2}+\epsilon)e_1-\sin(\frac{\alpha \pi}{2}+\epsilon)e_2\\
u_2 &=2.5e_2-e_3-5x_1z_1+5x_2 z_2+\sin(\frac{\alpha \pi}{2}+\epsilon)e_1-\cos(\frac{\alpha \pi}{2}+\epsilon)e_2\\
u_3 &=4x_1 y_1-4x_2y-2+0.2 x_2^2-0.2x_1^2,
\end{split} \label{5.14}
\end{equation}
where $\epsilon$ is any small positive real number.\\
Note that, the control terms $u_i$ depend on fractional order $\alpha$. Hence, for the given fractional order $\alpha$ the optimal strength of the controls is utilized in contrast with the conventional controls.\\
Now the error system (\ref{5.13}) becomes,
\begin{equation}
\begin{split}
{}_0^C\mathrm{D}_t^{\alpha}e_1 &=\cos(\frac{\alpha \pi}{2}+\epsilon)e_1+\sin(\frac{\alpha \pi}{2}+\epsilon)e_2\\
{}_0^C\mathrm{D}_t^{\alpha}e_2 &=-\sin(\frac{\alpha \pi}{2}+\epsilon)e_1+\cos(\frac{\alpha \pi}{2}+\epsilon)e_2\\
{}_0^C\mathrm{D}_t^{\alpha}e_2 &=-5.5e_3.
\end{split} \label{5.15}
\end{equation}
The eigenvalues of the coefficient matrix of linear system (\ref{5.15}) are,
\begin{equation}
\lambda_\pm=e^{\pm i(\frac{\alpha \pi}{2}+\epsilon)} \quad \mathrm{and} \quad \lambda_1=-5.5.
\end{equation}
Clearly, all the eigenvalues satisfy the condition,
$|arg(\lambda)|>\frac{\alpha \pi}{2}$.\\
So, the error system (\ref{5.15}) is stable and hence the systems (\ref{5.10}) and (\ref{5.11}) get synchronized.\\
 Figures \ref{Fig10}(a)-(c) show the synchronization where  the response system is shown by the dashed line. Here we have taken $\alpha=0.95$ and the initial conditions as $x_1(0) = -0.94$, $y_1(0)= 0.97$, $z_1(0) = -0.699$, $x_2(0) = -0.9$, $ 
 y_2(0) = -0.9$ and $z_2(0) = 0.6$. In Figure \ref{Fig10}(d), the error terms $e_i(t)$ of drive and response systems are plotted.

\section{Comment on a case of incommensurate order} \label{section7}
The behavior of the commensurate fractional order systems is relatively simpler than that of incommensurate order. Analyzing stability is a difficult task when the system is incommensurate.
\par If the commensurate order system is stable for some value $\alpha*$ then it remains stable for all the orders $\alpha\in(0,\alpha*)$. We cannot expect such nice behavior from the incommensurate order systems. We illustrate this with the following example.
\par
\begin{figure*}[h]
	\begin{tabular}{c c}
		\subfloat[$\beta=0.3$, $\alpha=\gamma=1$, system is unstable ]{\includegraphics[width=0.5\textwidth]{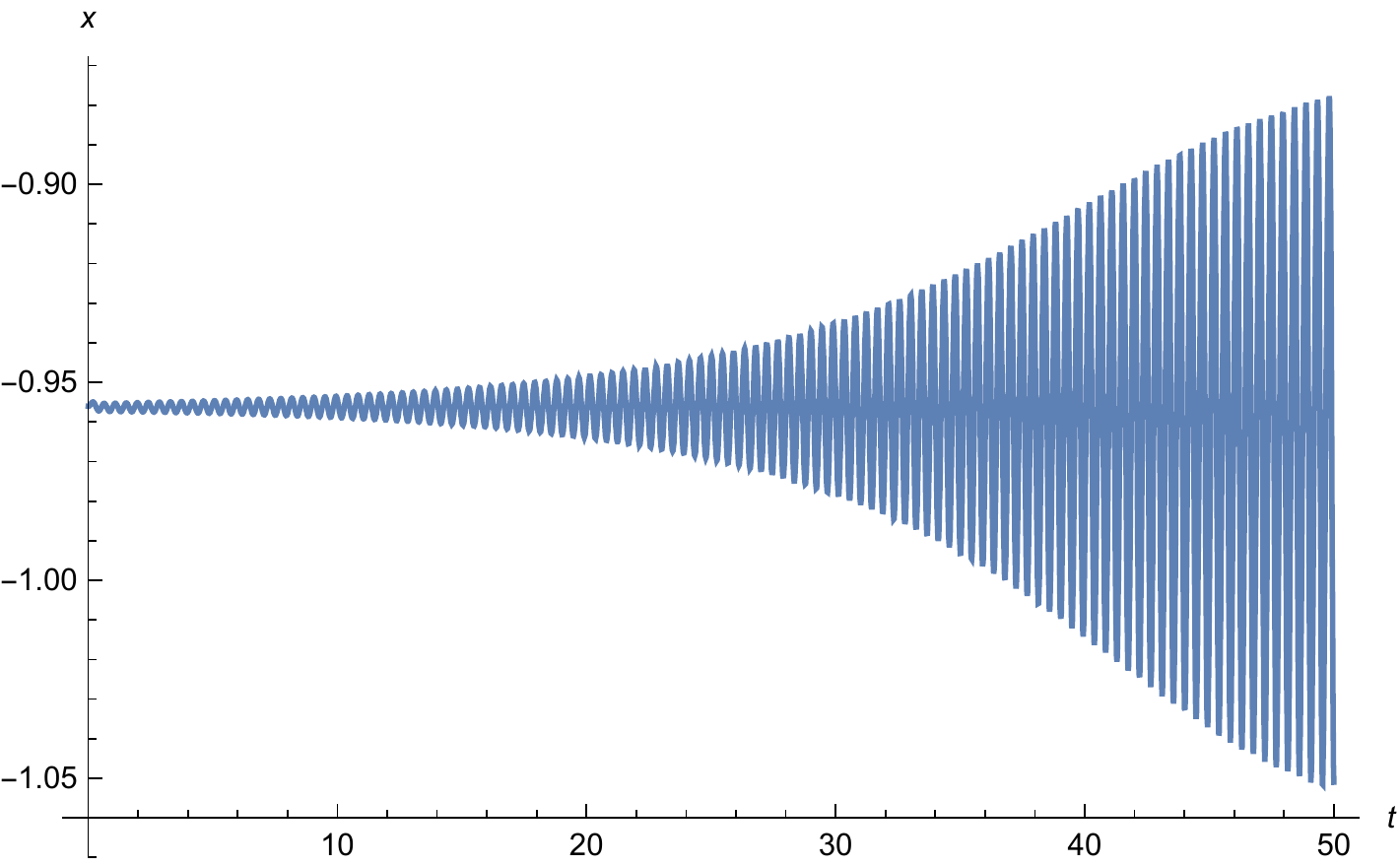}} 
		& 
		\subfloat[$\beta=0.4$, $\alpha=\gamma=1$, system is stable ]{\includegraphics[width=0.5\textwidth]{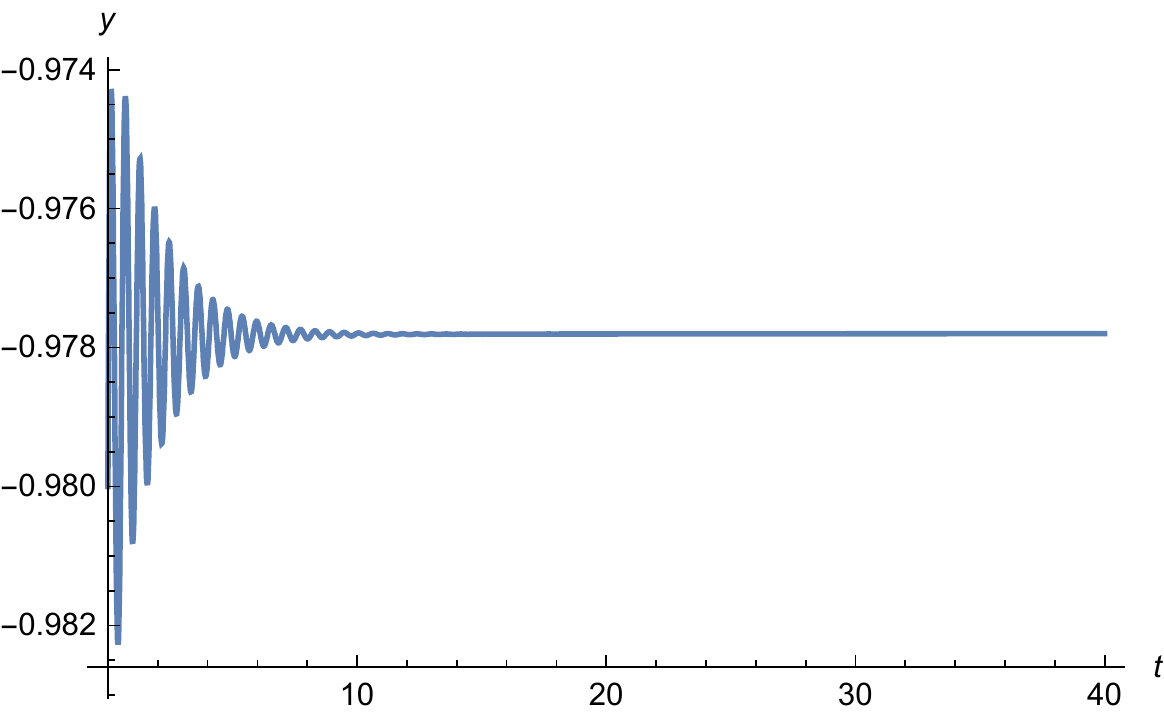}}
		\\
		\subfloat[  $\beta=0.92$, $\alpha=\gamma=1$, system is chaotic ]{\includegraphics[width=0.6\textwidth]{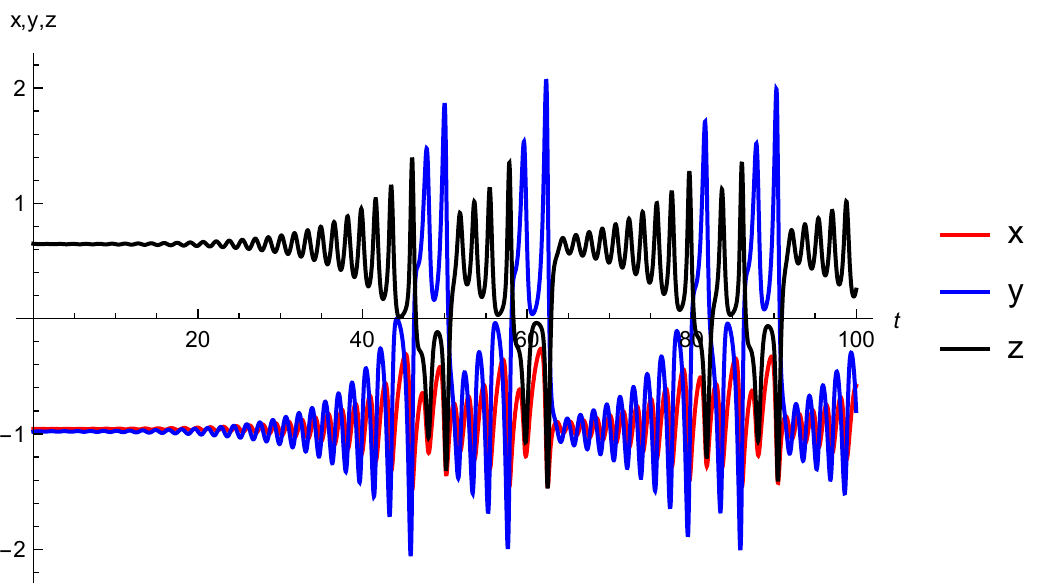}}
	\end{tabular}
	\caption{}
	\label{Fig11}
\end{figure*}

 Consider the system  (\ref{5.7}) with $\alpha=\gamma=1$. The system is chaotic for $\alpha=0.92$ and is stable for $\alpha=0.4$. However, it does not remain stable for all $\alpha\in(0,0.4)$ e.g. $\alpha=0.3$ produces unstable oscillations (cf. Fig. \ref{Fig11}).


\begin{table}[h]
	\begin{center} 
		\begin{tabular}{|c|c|c|c|c|c|}
			\hline
			Order of derivative & \multicolumn{3}{c|}{IMFOS evaluated at equilibrium points} & Stability & Corresponding figures\\
			\cline{2-4}
			 & ${\bm O}$ & ${\bm E_1}$ & ${\bm E_3}$ & & \\
			\hline
			$\beta=0.3$, $\alpha=\gamma=1$	&	$0.15708 >0$ & $0.014928 >0$ & $0.0184283 >0$ & Unstable & Figure \ref{Fig11}(a) \\
			\hline
			$\beta=0.4$, $\alpha=\gamma=1$	&	$0.314159 >0$ & $-0.00429731 <0$ & $0.00238612 >0$ & Stable & Figure \ref{Fig11}(b) \\
			\hline
			$\beta=0.92$, $\alpha=\gamma=1$	&	$0.0628319 >0$ & $0.00217138 >0$ & $0.00338347 >0$ & Chaotic & Figure \ref{Fig11}(c) \\
			\hline		
		\end{tabular}
		\caption{Weired behavior of incommensurate order system}
		\label{Tab3}
	\end{center}
\end{table} 

\section{Conclusion} \label{section8}
This article presents a new example of chaotic differential dynamical system. The bifurcation analysis with respect to various parameters is presented in the details. The stability analysis of fractional order generalization of the proposed system is described. Chaos in the commensurate as well as incommensurate order cases is investigated. The chaos in the system is controlled using a simple linear controller. The synchronization in the system is achieved using a nonlinear feedback controller. The peculiarity of this control is that it depends on fractional order. Unlike in other control strategies, the coupling strength is adjustable with the fractional order.

\section*{Acknowledgment}
S. Bhalekar acknowledges  the Science and Engineering Research Board (SERB), New Delhi, India for the Research Grant (Ref. MTR/2017/000068) under Mathematical Research Impact Centric Support (MATRICS) Scheme. M. Patil acknowledges Department of Science and Technology (DST), New Delhi, India for INSPIRE Fellowship (Code-IF170439). Authors are grateful to the anonymous reviewers for their insightful comments leading to the improved manuscript.

\end{document}